\documentclass[ALICE,manyauthors]{cernphprep}

\usepackage[pdftex]{hyperref}
\usepackage{amssymb}
\usepackage{cite}
\RequirePackage{graphicx}
\newcommand{\Pom}{$\mathbb{P}$}
\newcommand{\Reg}{$\mathbb{R}$}
\newcommand{\sigmainel}{\sigma_\mathrm{INEL}}
\newcommand{\sigmaSD}{\sigma_\mathrm{SD}}

\newcommand{\mbor}{$\mathrm{MB}_\mathrm{OR}$}
\newcommand{\mband}{$\mathrm{MB}_\mathrm{AND}$}

\begin{document}%
%
%
\begin{titlepage}
\PHnumber{2012-138}                 
\PHdate{27 August 2012}              
%
%
\title{Measurement of inelastic, single- and double-diffraction cross sections in proton--proton collisions at the LHC with ALICE}
\ShortTitle{Measurement of inelastic and diffractive cross sections}   
%
\Collaboration{ALICE Collaboration%
         \thanks{See Appendix~\ref{app:collab} for the list of collaboration
                      members}\\[1cm]
                      This publication is dedicated to the memory of our colleague A.B. Kaidalov who recently passed away.
                      }
\ShortAuthor{ALICE Collaboration}      
\begin{abstract}
Measurements of cross sections of inelastic and diffractive processes in
proton--proton collisions at LHC energies were carried out with the
ALICE detector. The fractions of diffractive processes in inelastic collisions were determined
from a study of gaps in charged particle pseudorapidity distributions:  for single diffraction (diffractive mass
$M_X < 200$ GeV/$c^2$)
$\sigma_{\rm SD}/\sigma_{\rm INEL} = 0.21 \pm 0.03, 0.20^{+0.07}_{-0.08}$,
and $0.20^{+0.04}_{-0.07}$, respectively at centre-of-mass energies $\sqrt{s} = 0.9, 2.76$, and 7~TeV;
for double diffraction (for a pseudorapidity gap $\Delta\eta > 3$)
$\sigma_{\rm DD}/\sigma_{\rm INEL} = 0.11 \pm 0.03, 0.12 \pm 0.05$,
and $0.12^{+0.05}_{-0.04}$, respectively at $\sqrt{s} = 0.9, 2.76$, and 7~TeV.
To measure the inelastic cross section, beam properties were determined with van der Meer scans, and, using a
simulation of
diffraction adjusted to data, the following values were obtained:
$\sigma_{\rm INEL} = 62.8^{+2.4}_{-4.0} (model) \pm 1.2 (lumi)$ mb at
$\sqrt{s} =$ 2.76~TeV
and $73.2^{+2.0}_{-4.6} (model) \pm 2.6 (lumi)$ mb at $\sqrt{s}$ = 7~TeV.
The single- and double-diffractive cross sections were calculated combining
relative rates of diffraction with inelastic cross sections.
The results are compared to previous measurements at proton--antiproton and
proton--proton colliders at lower energies, to measurements by other experiments at the
LHC, and to theoretical models.
\end{abstract}
\end{titlepage}
\setcounter{page}{2}
\section{Introduction}
\label{introduction}
The cross sections of inelastic and diffractive processes in proton--proton (pp)
collisions are among the basic observables used to characterize the global properties of
interactions, and thus are always a subject of interest at a new centre-of-mass energy. The behaviour of hadronic cross sections at high energies
 is usually described in the framework of  Regge theory~\cite{ReggeTheory} and its various QCD-inspired interpretations~\cite{ReggeQCD}.
As these collisions are dominated by relatively small-momentum transfer processes, such
measurements contribute to the theoretical understanding of QCD in the non-perturbative regime.
Recent developments in the field can be found in Refs. \cite{KMR,GLM,Ostapchenko,Goulianos,KP1}. As the LHC explores hadron collisions at centre-of-mass energies (up to $\sqrt{s}$ = 7~TeV used in the present analysis), corresponding to laboratory energies between $4 \times 10^{14}$ and $2.6 \times 10^{16}$~eV, close to the knee ($10^{15}$--$10^{16}$~eV) observed in the energy distribution of cosmic rays, these measurements are also relevant in this context.

It is customary to distinguish two contributions to the inelastic cross section:
diffractive processes and non-diffractive processes. At a centre-of-mass energy $\sqrt{s}$ = 1.8~TeV, at the
Tevatron, diffractive processes (single and double diffraction combined)
represent about 25\,$\%$ of inelastic collisions \cite{CDF}. At LHC energies, it is expected
that diffractive processes also account for a large fraction of the inelastic cross
section.

When presenting LHC measurements such as
particle momentum distributions, cross sections, etc. for Non-Single Diffractive (NSD) or Inelastic
(INEL) event classes, the uncertainty on the diffractive processes may dominate the overall
systematic error (see, for instance, Ref.~\cite{ALICE_EPJC68}). Therefore, it is essential
to measure, as precisely as possible, the properties of these processes. In addition, the nucleon--nucleon inelastic cross section is a basic parameter used as an input for model calculations
to determine the number of participating nucleons and the number of nucleon--nucleon binary collisions
for different centrality classes in heavy-ion collisions~\cite{GlaubModel}, the main focus of the ALICE scientific programme. This publication reports measurements of inelastic pp cross sections with a precision
better than 6$\%$, and emphasizes the importance of diffraction processes in such measurements.

The ALICE detector was used to measure the properties of gaps in the pseudorapidity distribution of particles emitted in pp collisions, in order to estimate the relative contributions of diffractive processes.
This publication is organized as follows: in Section~\ref{diffraction} we discuss diffractive processes and explain the definitions of diffraction adopted in this article; Section~\ref{experiment} gives a short description of the ALICE detector elements relevant
to this study, and describes the data samples used here and data-taking conditions; Section~\ref{measurement} presents relative rates of diffractive processes as measured from a pseudorapidity gap analysis, used to adjust these rates in the Monte Carlo event generators; Section~\ref{scans} discusses van der Meer beam scans, used to determine the LHC luminosity and the cross section corresponding to the trigger selection; in Section~\ref{crosssection} the simulation adjusted with our measurement is used to determine the inelastic cross section from the measured trigger cross section, and
in turn the cross sections for diffractive processes; finally a comparison is made between the ALICE cross section measurements and data from other experiments. The results are also compared with predictions from a number of models.

\begin{figure}[th!]
\includegraphics[width=.99\textwidth]{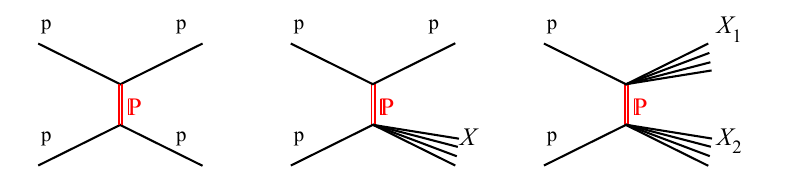}
\caption{
Lowest order Pomeron exchange graphs contributing to elastic (left), to single- (middle) and to double-diffractive (right) proton--proton scattering. \Pom\ stands for Pomeron, p for proton and $X$ ($X_1$, $X_2$) for the diffractive system(s).}
\label{Fig:fig1}
\end{figure}[th!]

\begin{figure}[ht]
\includegraphics[width=.99\textwidth]{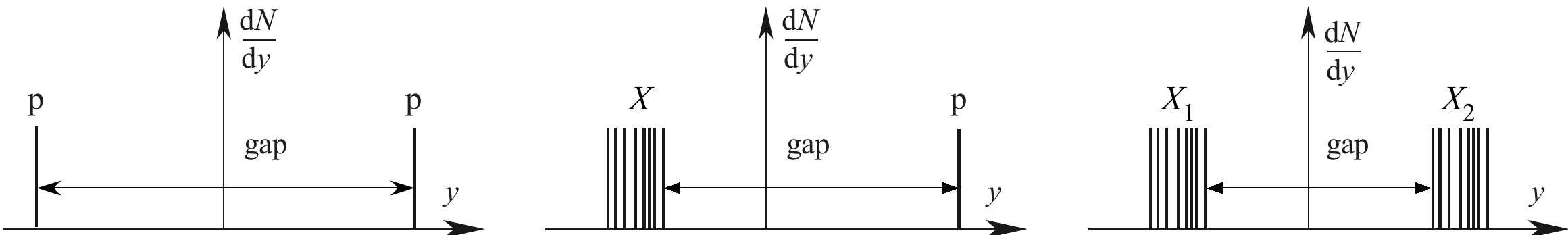}
\caption{Schematic rapidity ($y$) distribution of outgoing particles in elastic (left), in single- (middle),  and in double-diffraction (right) events,
showing the typical rapidity-gap topology.}
\label{Fig:fig2}
\end{figure}

\section{Diffraction}
\label{diffraction}
\subsection{Diffractive processes}
In  Regge theory at high energies, diffraction proceeds via the exchange of Pomerons
(see Ref.~\cite{ReggeTheory}). The Pomeron is a colour singlet object with the quantum
numbers of the vacuum, which dominates the elastic scattering amplitude at high energies.
The Feynman diagrams corresponding to one-Pomeron exchange in elastic, single- and double-diffraction
processes are shown in Fig.~\ref{Fig:fig1}. Single- and double-diffraction processes, ${\rm p} + {\rm p} \rightarrow {\rm p} + X$ and
${\rm p} + {\rm p} \rightarrow X_1 + X_2$, where $X$ ($X_1$, $X_2$) represent diffractive system(s), are closely related
to small-angle elastic scattering. These processes can be considered as binary collisions in which either or
both of the incoming protons may become an excited system, which decays into stable final-state particles.
Single Diffraction (SD) is similar to elastic scattering except that one of the protons breaks up, producing
particles in a limited rapidity region. In Double Diffraction (DD), both protons break up.

In SD processes, there is a rapidity gap between the outgoing proton and the other particles produced in the
fragmentation of the diffractive system of mass $M_X$ (Fig.~\ref{Fig:fig2} middle). For high masses, the average gap
width is $\Delta\eta \simeq \Delta y \simeq \ln(s/M_{X}^2) = - \ln \xi$, where $\xi = M_X^2/s$. Typically, at $\sqrt{s}$ = 7 TeV, $\Delta\eta$ varies from 13 to 7 for $M_X$ from 10 to 200~GeV/$c^2$.
In DD processes, there is a rapidity gap between the two diffractive systems (Fig.~\ref{Fig:fig2} right). The average
gap width in this case is $\Delta\eta \simeq \Delta y \simeq  \ln(ss_0/M_{X_1}^2M_{X_2}^2)$,
where the energy scale $s_0$ = 1~GeV$^2$, and $M_{X_1}$, $M_{X_2}$ are the diffractive-system masses. Typically, at $\sqrt{s}= 7$~TeV, one
expects $\Delta\eta \simeq 8.5$ for $M_{X_1} = M_{X_2} =$ 10~GeV/$c^2$.

Experimentally, there is no possibility to distinguish large rapidity gaps caused by Pomeron exchange from those caused by other colour-neutral exchanges (secondary Reggeons), the separation being model-dependent.
Therefore, in this study, diffraction is defined using a large rapidity gap as signature,
irrespective of the nature of the exchange. SD processes are those having a gap in rapidity from the leading proton limited by the value of the diffractive mass $M_X < 200$~GeV/$c^2$ on the other side ({\it i.e.} at $\sqrt{s}$ = 7~TeV, $\Delta\eta \gtrsim 7$); other inelastic events are considered as NSD. The choice of the upper $M_X$ limit corresponds approximately to the acceptance of our experiment and was used in previous measurements~\cite{UA5}. DD processes are defined as NSD events with a pseudorapidity gap
$\Delta \eta > 3$ for charged particles. The same value was used for separation between DD and Non-Diffractive (ND) processes  in another measurement~\cite{affolder}.

\subsection{Simulation of diffraction}
Diffraction processes are described by the distribution of the mass (or masses) of the
diffractive system(s), the scattering angle (or the four-momentum transfer $-t$), and the diffractive-system
fragmentation. The results depend only weakly on the assumption made for the distribution in $t$, and in all models, calculating acceptance and efficiency corrections, we integrated over this dependence. The $t$-distribution
and fragmentation of diffractive systems are simulated with the PYTHIA6
(Perugia-0, tune 320) \cite{PYTHIA6} and PHOJET \cite{PHOJET} Monte Carlo generators.
Both  PYTHIA6 and PHOJET  Monte Carlo generators give a reasonable description of UA4 data ~\cite{UA4sd} on charged particle pseudo-rapidity distribution in SD events.

\begin{figure}[th!]
\begin{center}
\includegraphics[width=.49\textwidth]{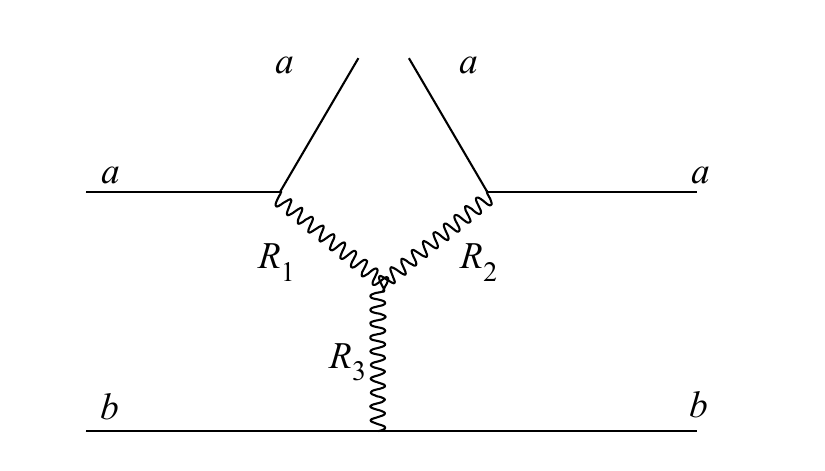}
\end{center}
\caption{
Triple-Reggeon Feynman diagram occurring in the calculation of the amplitude for single diffraction, corresponding to
the dissociation of hadron b in the interaction with hadron a. (See Ref.~\cite{ReggeTheory}). Each of the Reggeon legs
can be a Pomeron or a secondary Reggeon (e.g. $f$-trajectories), resulting in eight different combinations of Pomerons and
Reggeons. In the text, we use the notation $(R_1R_2)R_3$ for the configuration shown in this figure.}
\label{Fig:fig3}
\end{figure}

The main source of uncertainty in the simulation of diffractive collisions comes from the uncertainty on
the $M_X$ distribution (see, for example, the discussion in~\cite{OstapModel}). In Regge theory, in the single Regge pole
approximation, the SD cross section (${\rm d}\sigma/{\rm d}M_X$) for producing a high-mass system, $M_X$, is dominated by
the diagram shown in Fig.~\ref{Fig:fig3}. In the general case, each of the legs labeled
$(R_1R_2)R_3$, can be a Pomeron \Pom\ or a secondary Reggeon \Reg\ (e.g. the $f$-trajectory)~\cite{ReggeTheory}.
At very high energies, the SD process is dominated by the (\Pom\Pom)\Pom\ and (\Pom\Pom)\Reg\ terms, which have similar energy dependence, but a different $M_X$
dependence. The (\Pom\Pom)\Pom\ term is proportional to $1/M_X^{1+2\Delta}$ and the (\Pom\Pom)\Reg\ term to $1/M_X^{2+4\Delta}$, where $\Delta = \alpha_{\rm P}-1$, with
$\alpha_{\rm P}$ the intercept of the Pomeron trajectory. The (\Pom\Pom)\Reg\ term dominates the process
at very low mass, but vanishes at higher masses ($M_X^2 >> s_0$),
because the corresponding Regge trajectory has intercept smaller than unity.

\begin{figure}[th!]
\begin{center}
\includegraphics[width=.42\textwidth]{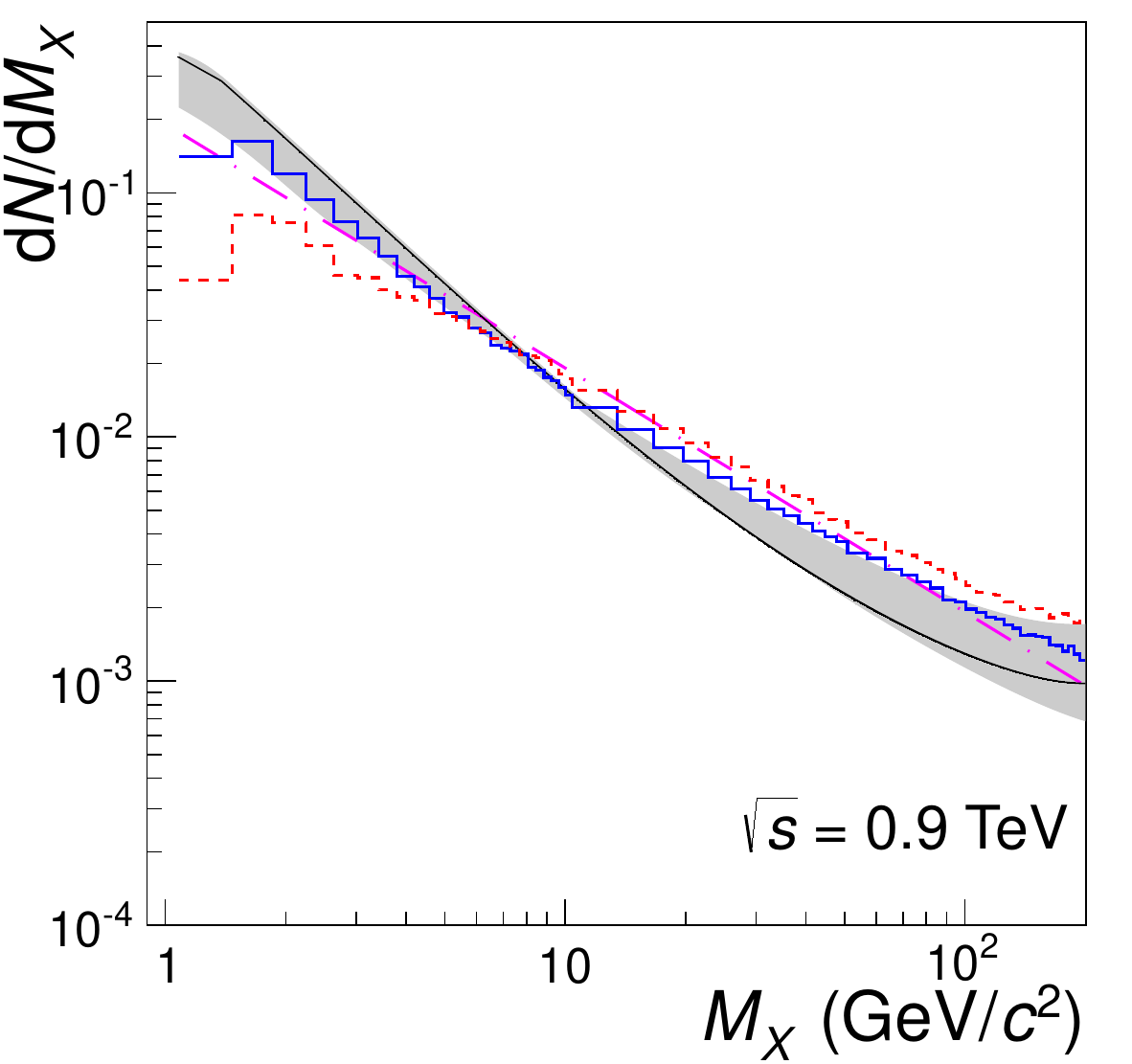}
\includegraphics[width=.42\textwidth]{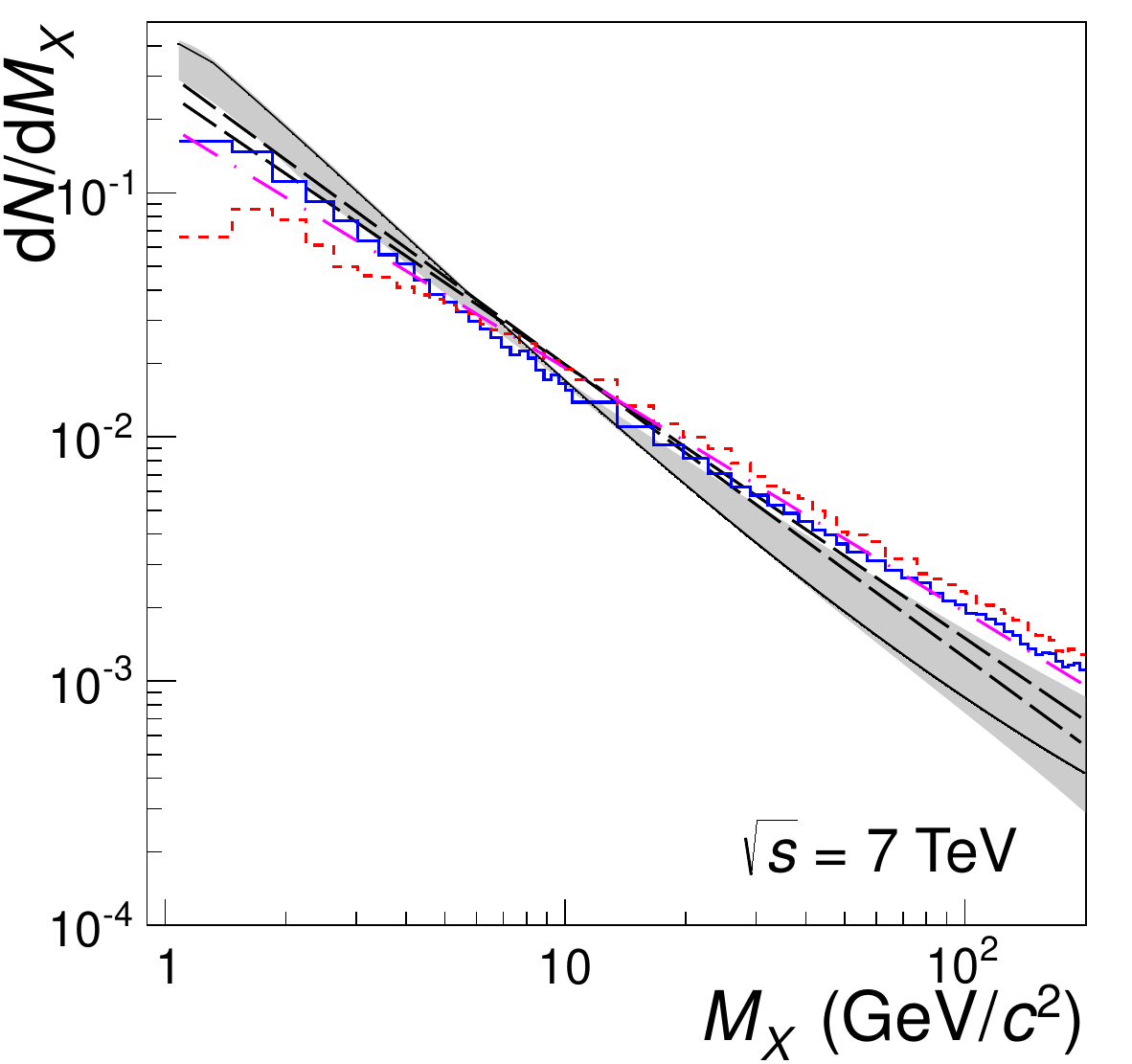}
\end{center}
\caption{
Diffractive-mass distributions, normalized to unity, for the SD process in pp collisions at $\sqrt{s}$ = 0.9~TeV (left) and $\sqrt{s}$ = 7~TeV (right), from Monte Carlo generators PYTHIA6 (blue histogram), PHOJET (red dashed-line histogram), and model~\cite{KP1}
(black line) --- used in this analysis for central-value estimate. The shaded area around the black line is delimited by (above at lower masses, below at higher masses) variation of the model~\cite{KP1}, multiplying the distribution by a linear function which increases the probability at the threshold mass by a factor 1.5 (keeping the value at upper-mass cut-off unchanged, and then renormalizing the distribution back to unity), and by (below at lower masses, above at higher masses) Donnachie--Landshoff parametrization~\cite{DL}. This represents the variation used for systematic-uncertainty estimates in the present analysis. A $1/M_X$ line is shown for comparison (magenta dotted-dashed line). At $\sqrt{s}$ = 7~TeV (right) black dashed-lines show $1/M_X^{1+2\Delta}$ distributions with $\Delta = 0.085$ and $0.1$ also used with PYTHIA8 event generator in the ATLAS measurement of inelastic cross section ~\cite{ATLAS_InelXS}.
}\label{Fig:fig4}
\end{figure}

In both the PYTHIA6 and PHOJET generators, the diffractive-mass distribution for the SD processes
is close to $1/M_X$ (Fig.~\ref{Fig:fig4}), which corresponds to the (\Pom\Pom)\Pom\ term with $\Delta = 0$. However, experimental data show that at low masses the dependence is steeper than $1/M_X$.
This is discussed, for example, in publications by
the CDF collaboration~\cite{CDF}, and supports the values of $\Delta > 0$ and also the
above theoretical argument for inclusion of terms other than (\Pom\Pom)\Pom.
A recent version of PYTHIA having a steeper $M_X$ dependence at low masses, PYTHIA8 \cite{PYTHIA8}, uses an approximation with a $1/M_X^{1+2\Delta}$ dependence with $\Delta = 0.085$, based on the (\Pom\Pom)\Pom\ term in the Donnachie--Landshoff model~\cite{DL}.

For this study the $M_X$ distributions in PYTHIA6 and PHOJET were modified
so as to use
the distributions from model \cite{KP1} (Fig.~\ref{Fig:fig4}), which includes in the calculation of the SD cross section all eight terms contributing to the diagram of Fig.~\ref{Fig:fig3}. Their relative contributions are determined from a fit to  lower-energy data. The predictions of this model for the total, elastic, and diffractive cross sections at LHC energies ~\cite{KP2} are 
found to be consistent with measurements~\cite{ATLAS_InelXS,CMS_InelXS,Totem}.
The modification of PYTHIA6 and PHOJET consists in reproducing the model $M_X$ distribution, by 
applying weights to the generated events. Numerical values of the diffractive-mass distributions 
from this model, at the three centre-of-mass energies relevant to this publication, can be found in~\cite{KP3}.

In addition, the fractions of diffractive processes in the models were adjusted according to measurements presented here, by a normalization factor. In what follows, ``adjusted''
PYTHIA6 or PHOJET means that these event generators are used with the modified diffractive-mass distribution, and the modified
relative rate of diffractive processes.

In order to estimate the systematic errors coming from the uncertainty in the functional shape of the $M_X$ dependence, the following modifications were used: the model distribution was multiplied by a linear function $a M_X + b$, which is equal to unity at the upper diffractive-mass value $M_X =$ 200~GeV/$c^2$ and is equal to 0.5 or 1.5 at the diffractive-mass threshold, {\it i.e.} $M_X \approx 1.08$~GeV/$c^2$ (sum of proton and pion masses). 
A linear function was chosen because it is the simplest way to vary the relative fractions of low-mass (or non detected) and high-mass (or detected) single diffractive events.
The resulting variation is illustrated in~Fig.~\ref{Fig:fig4}, where the diffractive-mass distributions are normalized to have the integral between threshold and $M_X =$ 200~GeV/$c^2$ equal to unity. The influence of the change of the $M_X$ dependence on the SD rate is given roughly by the variation of the yield in the high-$M_X$ region (above $\simeq 10$~GeV/$c^2$, where the events are measured) relative to that in the low-$M_X$ region (where an extrapolation has to be used). The distribution from the Donnachie--Landshoff model~\cite{DL} was also used in the evaluation of the systematic uncertainties due to the extrapolation to low-$M_X$ region. The ATLAS collaboration, in their measurement of the inelastic cross section~\cite{ATLAS_InelXS}, used unmodified PYTHIA6 and PHOJET event generators, with an approximate $1/M_X$ dependence, the (\Pom\Pom)\Pom\ term of the Donnachie--Landshoff
model (as parameterized in PYTHIA8), around which they varied the mass distribution (see Fig.~\ref{Fig:fig4}), and also the calculations with model~\cite{KMR}.
Thanks to the collaboration of the authors of Refs. \cite{KMR}, \cite{GLM} and \cite{Goulianos} we were able to check that the single-diffraction mass dependencies of the corresponding models, when normalized at the $M_X$ = 200 GeV/$c^2$, are well within the limits assumed in this analysis.

 Concerning the simulation of DD processes in PYTHIA6 and PHOJET event generators, only their overall fraction was adjusted according to our data, otherwise it was left unmodified. However, all NSD events with pseudorapidity gap $\Delta \eta > 3$, including those flagged by a generator as ND, are considered to be DD. This way, processes with secondary Reggeon legs in a diagram analogous to that in Fig.~\ref{Fig:fig3} are also taken into account, albeit in a very model-dependent way. Therefore, the results for DD fractions and cross sections are subject to larger uncertainties than those for SD.

\section{Experiment description}
\label{experiment}
\subsection{The ALICE detector}
The ALICE detector is described in Ref.~\cite{ALICE_JINST}. The analysis presented here is mainly based on data from the
VZERO detector,  the Silicon Pixel Detector (SPD) and the Forward Multiplicity Detector (FMD). The SPD and the VZERO
hodoscopes provide trigger information for the selection of minimum-bias events and for van der Meer~\cite{vdM}
proton-beam scans. The Time-Projection Chamber (TPC) \cite{ALICE_TPC}
and the whole Inner Tracking System (ITS)  \cite{ALICE_ITS}, both situated in the ALICE
central barrel, are used in this study only to provide the interaction vertex position, from reconstructed tracks.

Throughout this publication,
the detector side at negative (positive) pseudorapidity is referred to as left or ``L-side'' (right or ``R-side''). The asymmetric arrangement of the detectors comes about because of the space constraints imposed by the ALICE muon arm on the L-side.

The two VZERO hodoscopes, with 32 scintillator tiles each, are placed on each side of the interaction region at $z \simeq 3.3$ m
(V0-R) and $z \simeq -0.9$ m (V0-L), covering the pseudorapidity ranges $2.8 < \eta < 5.1$ and $-3.7 < \eta < -1.7$, respectively
($z$ is the coordinate along the beam line, with its origin at the centre of the ALICE barrel detectors, oriented in the direction
opposite to the muon arm~\cite{ALICE_JINST}). Each hodoscope is segmented  into eight equal azimuthal angle $\varphi$ sectors and four equal  pseudorapidity $\eta$ rings. This implies that the pseudorapidity resolution is similar to that required for the binning ($\delta\eta = 0.5$) used for the analysis.
The time resolution of each hodoscope is better than 0.5~ns.

The SPD makes up the two innermost layers of the ALICE Inner Tracking System (ITS) and it covers the pseudorapidity ranges
$|\eta| < 2$ and $|\eta| < 1.4$, for the inner and outer layers respectively. The SPD has in total about $10^7$ sensitive pixels on 120 silicon ladders which were aligned using pp collision
data to a precision of 8~$\mu$m. The SPD can also be used to  provide the position of the interaction vertex by correlating hits in the two
silicon-pixel layers to obtain track elements called tracklets. The resolution achieved on the position of the vertex from the SPD is
slightly worse than that for the vertex from tracks reconstructed with the TPC and the whole ITS. It depends on the track multiplicity and is approximately 0.1--0.3~mm
in the longitudinal direction and 0.2--0.5~mm in the direction transverse to the beam line. If the vertex from reconstructed tracks is not
available, 
the vertex from the SPD is used.

The FMD consists of Si-strip sensors with a total of above $5 \times 10^4$ active detection elements, arranged in five rings perpendicular to the
beam direction, covering the pseudorapidity ranges $-3.4 < \eta < -1.7$ (FMD-3) and $1.7 < \eta  < 5.1$ (FMD-1 and FMD-2). Combining VZERO, SPD and FMD, ALICE has a continuous acceptance over a pseudorapidity interval of 8.8 units. 
\subsection{Event samples and data-taking conditions}
ALICE data were collected at three centre-of-mass energies ($\sqrt{s}$ = 0.9, 2.76, and 7 TeV), at low beam current and
low luminosity, hence corrections for beam backgrounds and event pileup in a given bunch crossing are small. The maximum average number of collisions per bunch crossing was 0.1 at $\sqrt{s}$ = 7~TeV.

The minimum-bias data used for the diffractive study were collected using the trigger condition \mbor, which requires
at least one hit in the SPD or in either of the VZERO arrays. This condition is satisfied by the passage of a
charged particle anywhere in the 8 units of pseudorapidity covered by these detectors.

For the van der Meer scan measurements, the trigger requirement was a time coincidence between hits in the two
VZERO scintillator arrays, V0-L and V0-R, \mband.

Control triggers taken for various combinations of filled and empty bunch buckets were used to measure beam-induced
background and accidental triggers due to electronic noise or cosmic showers. 
Beam backgrounds were removed offline using two conditions. First, VZERO counter timing signals, if present, had to be compatible with particles produced in collisions. Second, the ratio of the number of SPD clusters to the number of SPD tracklets is much higher in background events than in beam--beam collisions, thus a cut on this ratio was applied. The latter condition was adjusted using beam-background events selected with VZERO detector. The
remaining background fraction in the sample was estimated from the number of control-trigger events that passed the
event selection. It was found to be negligible for the three centre-of-mass energies, except in the case of the van der Meer scan at
$\sqrt{s}$ = 2.76 TeV at large displacements of the beams, as discussed in Section~\ref{scans}.

At each centre-of-mass energy, several data-taking runs were used, with different event pileup rates, in order to correct for
pileup, as described below. For the measurement of the inelastic cross sections, runs were chosen to be as close in
time as possible to the runs used for the van der Meer scans in order to ensure that the detector configuration had
not changed.

At $\sqrt{s}$ = 0.9~TeV, $7 \times 10^6$ events collected in May 2010 were used for diffractive studies. No van der Meer scan was performed
at this energy, hence the inelastic cross section was not measured by ALICE.

At $\sqrt{s}$ = 7~TeV, $75 \times 10^6$ events were used for diffractive studies, and van der Meer scans were performed five months apart during the
pp data-taking period (scan I in May 2010, scan II in October 2010).

The data at $\sqrt{s}$ = 2.76 TeV were recorded in March 2011, at an energy chosen to match the nucleon--nucleon centre-of-mass
energy in Pb--Pb collisions collected in December 2010. For diffraction studies, $23 \times 10^6$ events were used. One van der Meer scan was performed
(scan III in March 2011). Because of a technical problem the FMD was not used in diffraction measurements, resulting
in a larger systematic uncertainty in diffractive cross-section measurements at this energy.

\section{Measurement of relative rates of diffractive processes}
\label{measurement}
\subsection{Pseudorapidity gap study}
For this study the events were selected by the hardware trigger
\mbor\ followed by the ALICE offline event selection described in Section~\ref{experiment}. The pseudorapidity distribution of particles emitted in the collision is studied by associating the event vertex with a ``pseudo-track'' made from a hit in a cell of the SPD, of the VZERO or of the FMD detector. In the case of VZERO, the cells are quite large ($\delta\eta$ about 0.5), so for this detector hits were distributed randomly within the cell pseudorapidity coverage. \\
Note that the effective transverse-momentum threshold for the pseudo-track detection is very low (about 20 MeV/$c$).
It is practically pseudo-rapidity independent and determined by the energy loss in the material. This implies that our detector misses only a very small fraction of particles.

The vertex is reconstructed using information from the ITS and TPC, if possible. If it is not possible to form a vertex in this way, a position is calculated using the SPD alone. If this is also not possible (as it occurs in 10\% of cases), then a vertex is generated randomly using the measured vertex distribution. Such cases occur mainly where there is no track in the SPD and hit information is in the VZERO or FMD detectors only.

In the analysis described below, we separate the events into three categories, called ``1-arm-L'', ``1-arm-R'' and ``2-arm''. The purpose of the classification is to increase the sensitivity to diffractive processes. As will be described below, the categories 1-arm-L and 1-arm-R have an enriched single-diffraction component, while a subset of the 2-arm category can be linked to double diffraction.

We distinguish between ``one-track'' events and those having more than one track, {\it i.e.} ``multiple-track'' events. Let $\eta_{\rm L}$, $\eta_{\rm R}$ be the pseudorapidities  of the leftmost (lowest-pseudorapidity) and rightmost (highest-pseudorapidity) pseudo-tracks in the distribution respectively. If an event has just one pseudo-track, then $\eta_{\rm L} = \eta_{\rm R}$. We classify as one-track events all events satisfying the condition $\eta_{\rm R}-\eta_{\rm L}<0.5$ and having all pseudo-tracks within $45^{\circ}$ in azimuthal angle $\varphi$. For such events, we determine the centre of the pseudorapidity distribution as $\eta_{\rm C} = \frac{1}{2}(\eta_{\rm L}+\eta_{\rm R})$, and


\newcounter{romancounter}
\begin{list}
{({\it \roman{romancounter}}$\thinspace$)}{\usecounter{romancounter}
\setlength{\rightmargin}{\leftmargin}}
\item if $\eta_{\rm C}<0$ the event is classified as 1-arm-L;
\item if $\eta_{\rm C}>0$ the event is classified as 1-arm-R.
\end{list}

\begin{figure}
\begin{center}
\includegraphics[width=.47\textwidth]{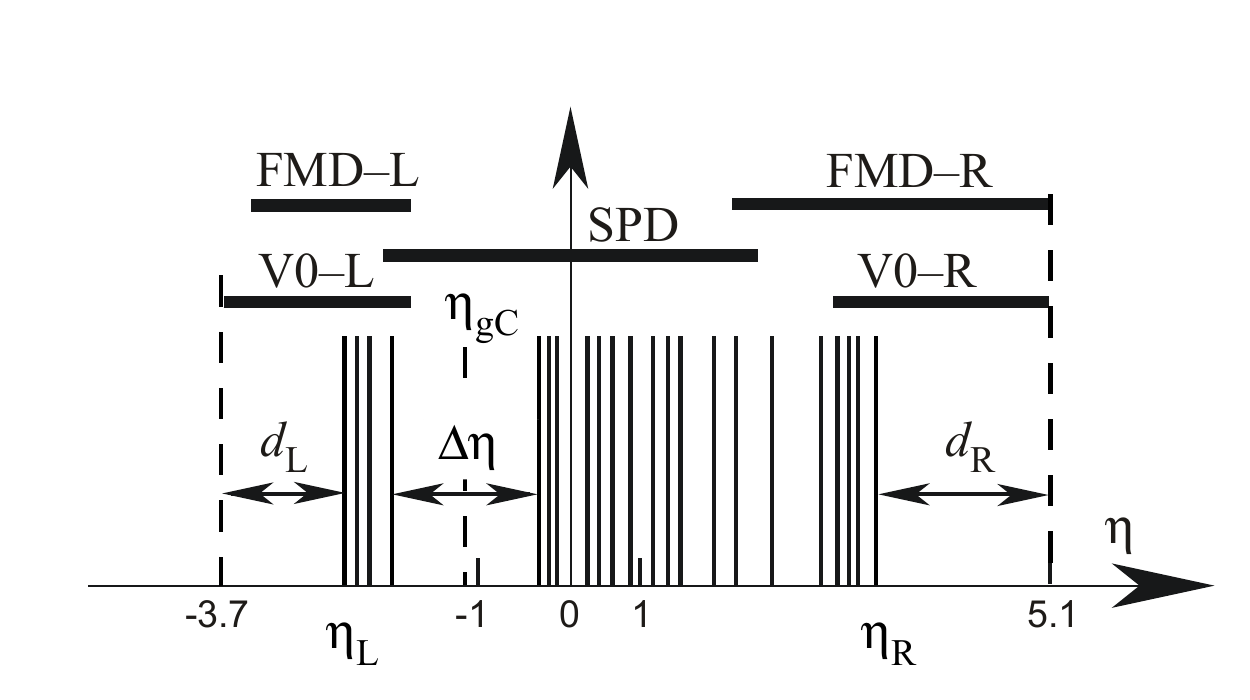}
\end{center}
\caption{
Pseudorapidity ranges covered by FMD, SPD and VZERO (V0-L and V0-R) detectors, with an illustration of  the distances
$d_{\rm L}$ and $d_{\rm R}$ from  the edges ($\eta_{\mathrm{L}}$ and $\eta_{\mathrm{R}}$, respectively) of the particle pseudorapidity distribution to the edges of the ALICE
detector acceptance (vertical dashed lines --- for the nominal interaction point position) and the largest gap $\Delta\eta$ between adjacent tracks. The centre of the largest gap is denoted $\eta_{\rm gC}$. L and R stand for Left and Right, respectively, following the convention defined in Section~\ref{experiment}.
}\label{Fig:fig6}
\end{figure}

The multi-track events are classified in a different way. For these events, we use the distance $d_{\rm L}$ from the track with pseudorapidity $\eta_{\rm L}$ to the lower edge of the acceptance, the distance $d_{\rm R}$ from the track with pseudorapidity $\eta_{\rm R}$ to the upper edge of the acceptance, and the largest gap $\Delta\eta$ between adjacent tracks (see Fig.~\ref{Fig:fig6}). Then,

\begin{list}
{({\it \roman{romancounter}}$\thinspace$)}{\usecounter{romancounter}
\setlength{\rightmargin}{\leftmargin}}
\item if the largest gap $\Delta\eta$ between adjacent tracks is larger than both $d_{\rm L}$ and $d_{\rm R}$, the event is classified as 2-arm;
\item if both of the edges $\eta_{\rm L}$, $\eta_{\rm R}$ of the pseudo-rapidity distribution are in the interval $-1\le\eta\le 1$, the event is classified as 2-arm;
\item otherwise, if $\eta_{\rm R}<1$ the event is classified as 1-arm-L, or if $\eta_{\rm L}>-1$ the event is classified as 1-arm-R;
\item any remaining events are classified as 2-arm.
\end{list}

\begin{figure*}
\begin{center}
\includegraphics[width=.42\textwidth]{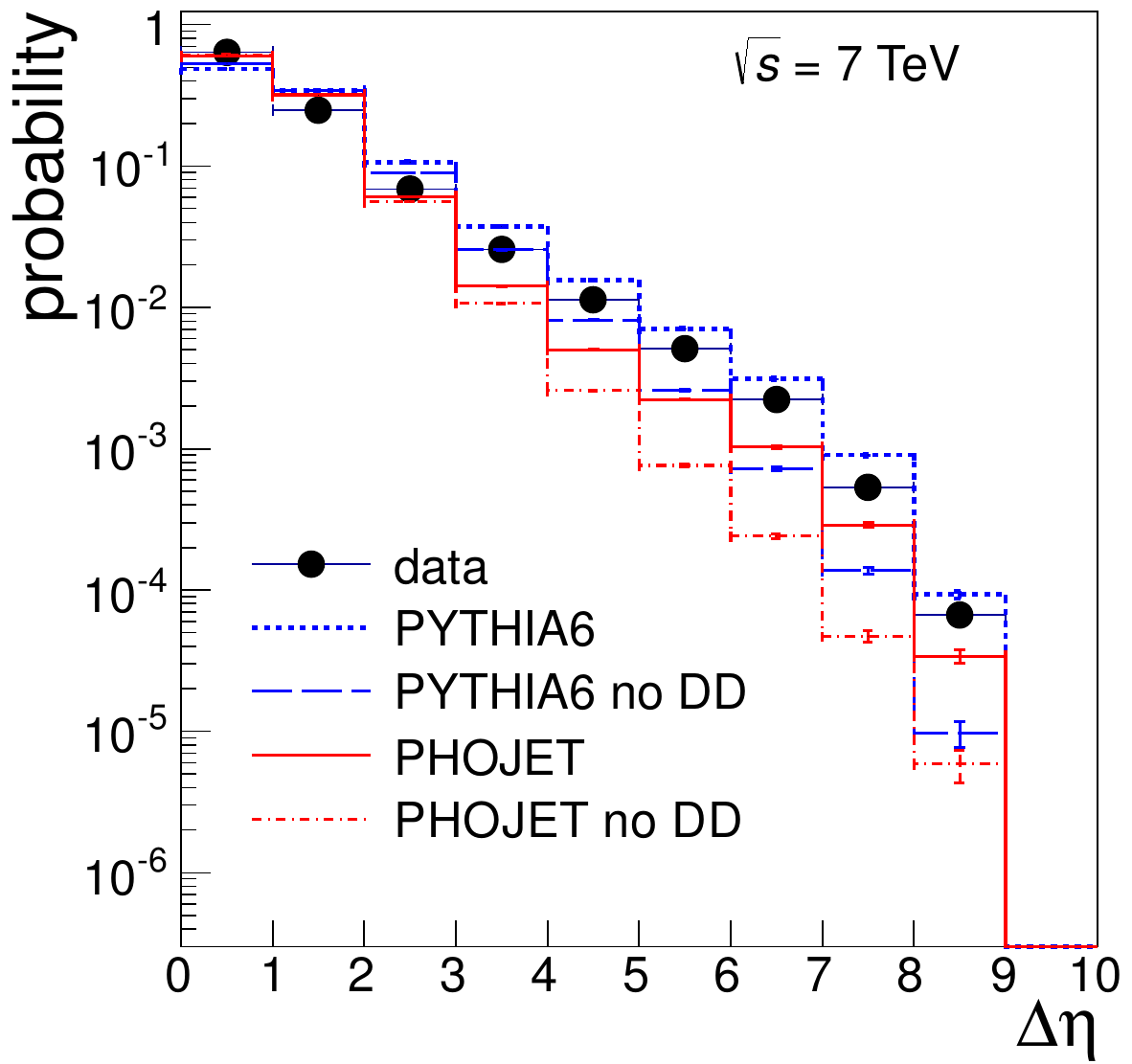}
\includegraphics[width=.42\textwidth]{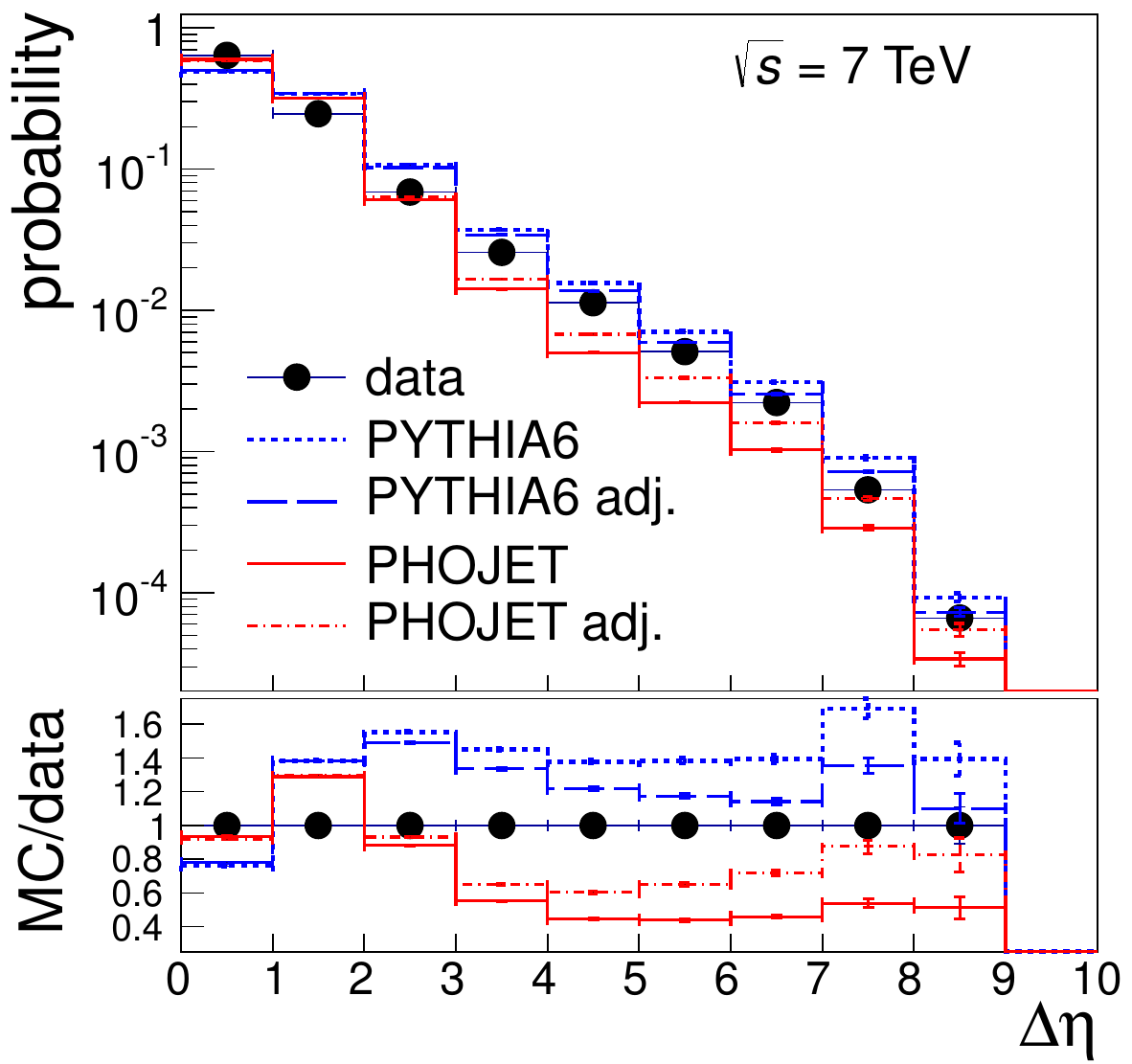}
\end{center}
\caption{
Largest pseudorapidity gap width distribution for 2-arm 
events, comparison between the data (black points) and various simulations (stage $d$). Left: dotted blue and solid red lines were obtained from default PYTHIA6 and PHOJET, respectively;
dashed blue and dashed-dotted red lines were obtained by setting the DD fraction to zero in PYTHIA6 and PHOJET, respectively. Right: dotted blue and solid red lines are the same as on the left side; dashed blue and dashed-dotted red lines are for adjusted PYTHIA6 and PHOJET, respectively; the ratio of simulation to data is shown below with the same line styles for the four Monte Carlo calculations.
}\label{Fig:fig8}
\end{figure*}

\begin{figure}
\begin{center}
\includegraphics[width=.42\textwidth]{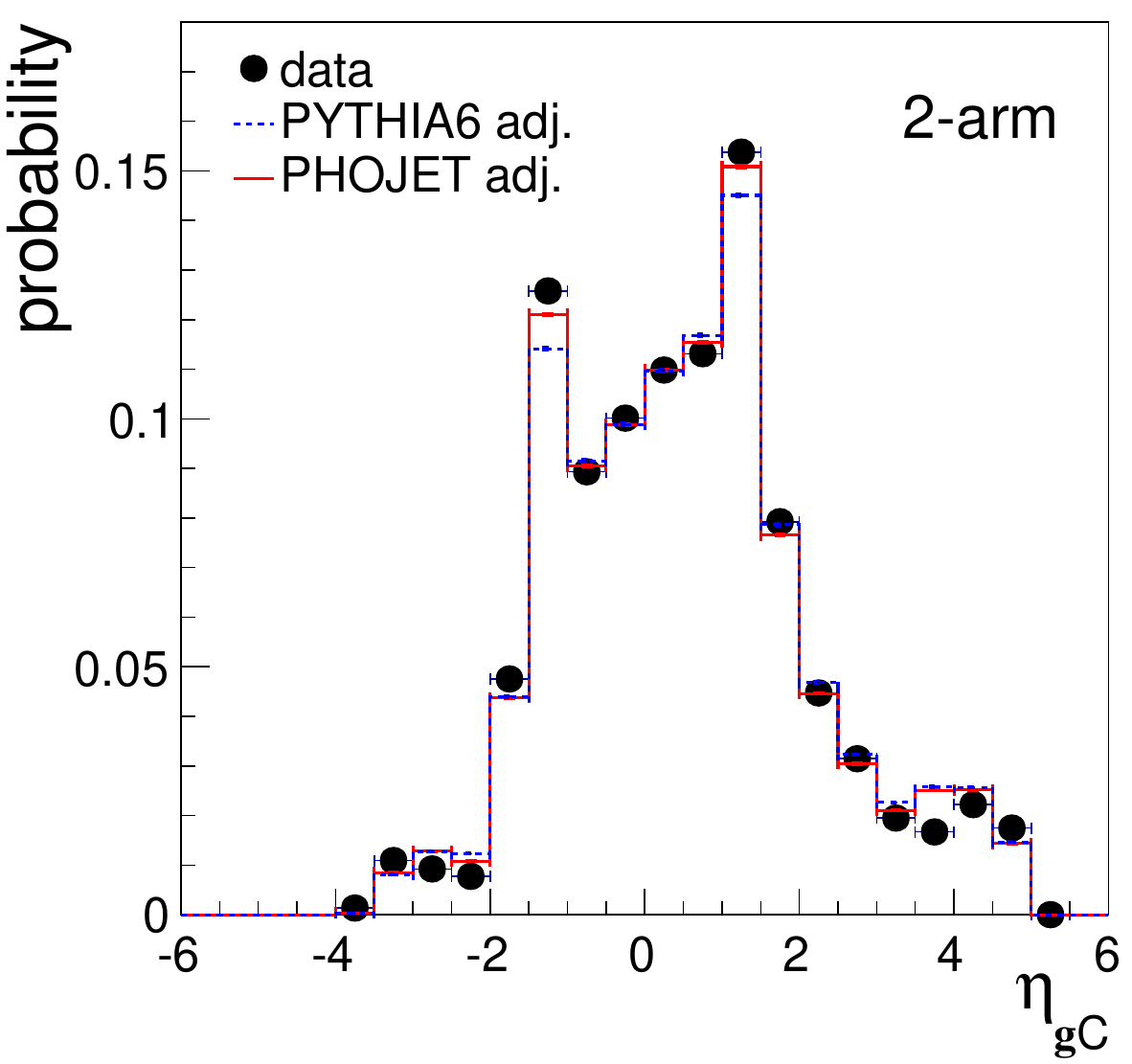}\\
\includegraphics[width=.42\textwidth]{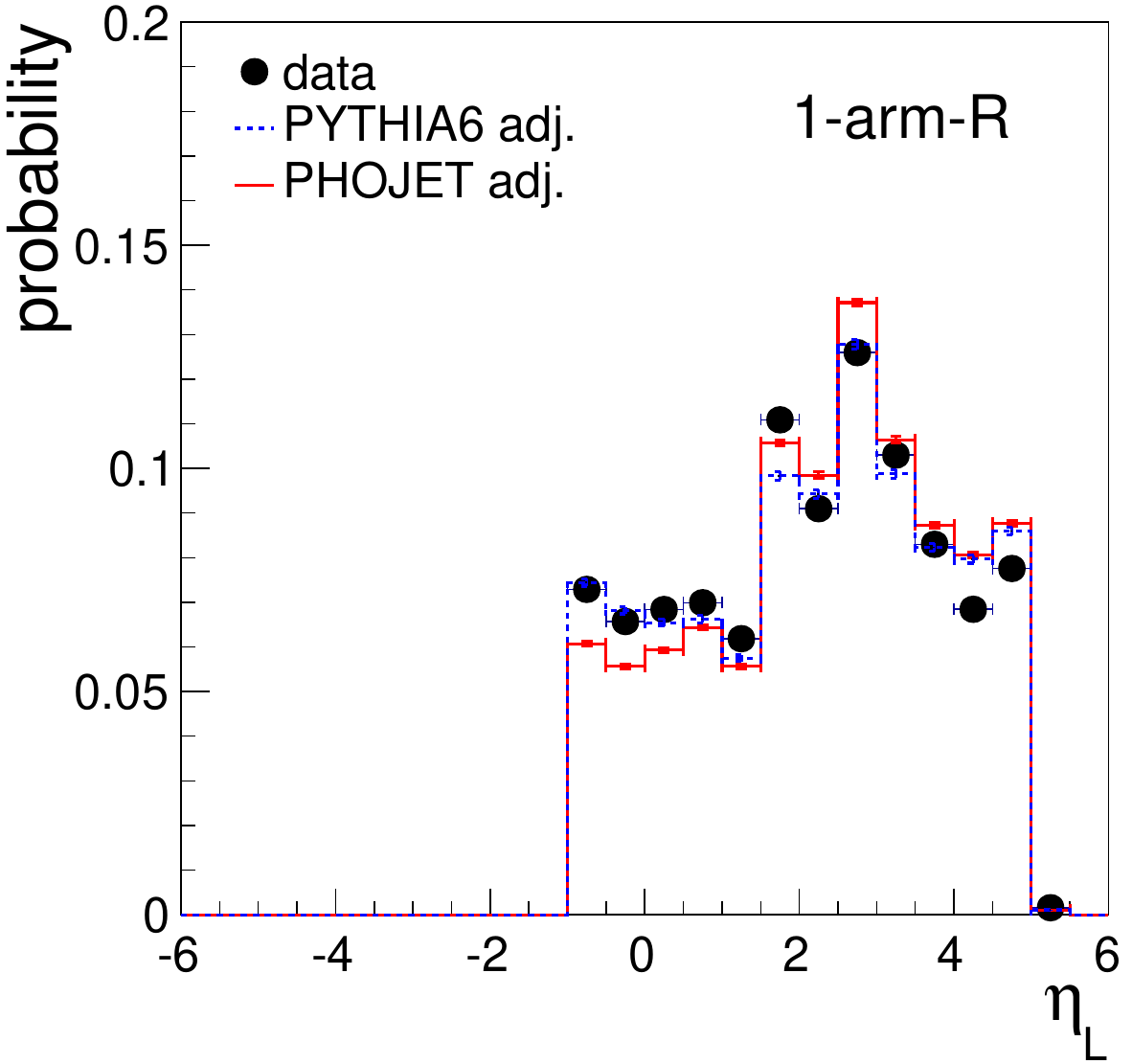}\\
\includegraphics[width=.42\textwidth]{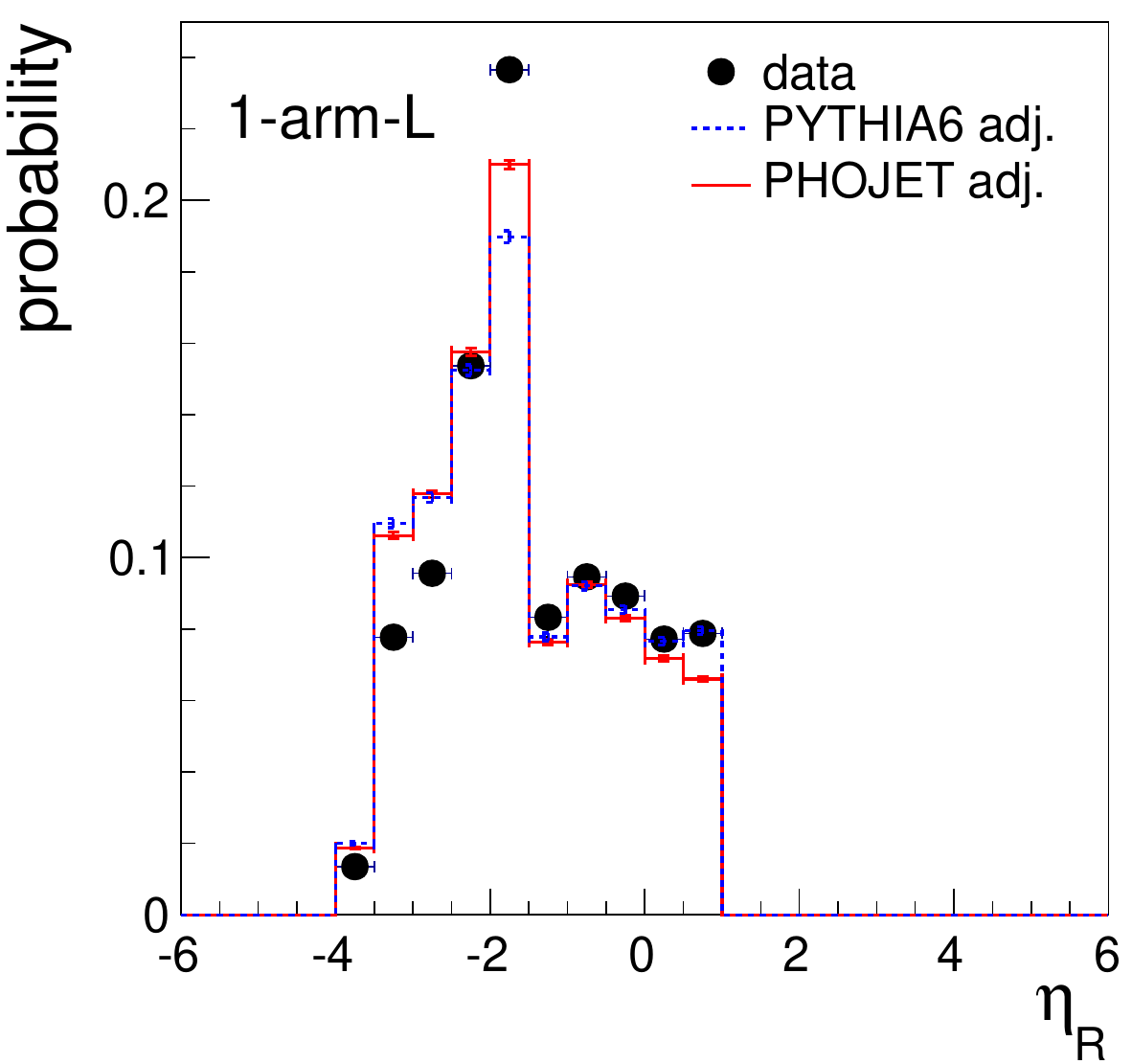}
\end{center}
\caption{
Comparison of reconstructed data versus adjusted Monte Carlo simulations (stage $d$), at $\sqrt{s}$ = 7 TeV. For 2-arm  event class (top),
pseudorapidity distributions of centre position ($\eta_{\rm gC}$) of the largest pseudorapidity gap;
distribution for 1-arm-L (middle) and 1-arm-R (bottom) event classes, respectively of the pseudorapidity of the right edge ($\eta_{\rm R}$)
and left edge ($\eta_{\rm L}$) of the pseudorapidity distribution.
}\label{Fig:fig7}
\end{figure}

The ALICE Monte Carlo simulation consists of four main stages: ($a$) event generation; ($b$) transport through material;
($c$) detector simulation, and ($d$) event reconstruction. In Figs.~\ref{Fig:fig8} and \ref{Fig:fig7}, we compare gap properties between
data and Monte Carlo simulation after event reconstruction (stage $d$).

In Fig.~\ref{Fig:fig8} left, the gap width distribution for 2-arm events is compared to simulations with and without DD, to illustrate the sensitivity to the DD fraction.
The gap width distribution at large $\Delta\eta$ cannot be described  by simulations without DD. However, the default DD fraction in PYTHIA6 significantly overestimates the distribution of large pseudorapidity gaps, while the default DD distribution in PHOJET significantly underestimates it. Adjustments to these fractions can bring the predictions of the two generators into better agreement with the data, and lead to a method to estimate the DD fraction. A similar approach was employed by the CDF collaboration~\cite{affolder}. The DD fractions in PYTHIA6 and PHOJET were varied in steps so as to approach the measured distribution.

The aim of the adjustment is to bracket the data. At the end of the adjustment the PYTHIA6 data still overestimate the data, and the PHOJET data underestimate it, but the agreement between data and Monte Carlo is brought to 10\% for the bin in closest agreement above $\Delta\eta = 3$ (see Fig.~\ref{Fig:fig8} right). Further adjustment leads to a deterioration in the shape of the $\Delta\eta$ distribution. The mean value between the PYTHIA6 and PHOJET estimates is taken as the best estimate for the DD fraction, and the spread between the two contributions, integrated over $\Delta\eta > 3$, is taken as a a measure of the model error.

Once the value for the DD fraction has been chosen, and its associated error estimated as described above, the measured 1-arm-L(R) to 2-arm ratios, which have negligible statistical errors, can be used to compute the SD fractions and their corresponding errors. For this purpose the efficiencies for the SD and NSD events to be detected as 1-arm-L(R) or 2-arm classes have to be known. The determination of these efficiencies is described later in this Section. A similar method was used by the UA5 collaboration in their measurement of diffraction~\cite{UA5}. In practice, we handle the L-side and R-side independently and
the SD fractions are determined separately for L-side and R-side single diffraction.

 In summary, three constraints from our measurements, the two 1-arm-L(R) to 2-arm ratios and the additional constraint obtained from the gap width distribution (Fig.~\ref{Fig:fig8} right), are used to compute the three fractions for DD events, L-side SD, and R-side SD events. The sum of the two latter values is then used to estimate the SD fraction of the overall inelastic cross-section. This way the Monte Carlo event generators PYTHIA6 and PHOJET are adjusted using the experimental data, and this procedure is repeated for different assumptions about the diffractive-mass distribution for SD processes, as discussed in Section~\ref{diffraction}.



For the $\sqrt{s}$ = 2.76 TeV run, the FMD was not used in the analysis, resulting in a gap in the detector acceptance, so the fraction of DD events in the
Monte Carlo
generators was not adjusted using the gap distribution for this energy. The resulting DD fraction of the inelastic cross section, however, was modified due to the adjustment of the SD fraction and the experimental definition of DD events. This results in a larger systematic error on the measured DD cross section
at this energy.

In Fig.~\ref{Fig:fig7} we compare other pseudorapidity distribution properties after event generator adjustment.
In addition to the quantities defined above, we use in this comparison the centre position $\eta_{\rm gC}$ of the $\Delta\eta$ pseudorapidity gap.
The observed basic features of the edges of pseudorapidity distributions and gaps are reasonably well reproduced by the adjusted simulations for
$|\eta| \geq 1.5$, and more accurately for $|\eta| \leq 1.5$.
Fig.~\ref{Fig:fig7}
shows the $\sqrt{s}$ = 7~TeV case for illustration. The agreement
between data and simulation is similar at $\sqrt{s}$ = 0.9~TeV and 2.76~TeV.\\
%
We note that there is less material in the R-side of the ALICE detector. With the adjusted Monte-Carlo generators we have obtained a good description for the 1-arm-R event class. On the L-side, there is more material between V0-L and the interaction point and the distribution of material is not as precisely known as on the other side. For this reason we have used a larger error margin in  our study of the corresponding systematic uncertainty.\\
Several tests were made to ensure that the material budget and the properties of the detectors do not modify the correlations between observables and rates of diffractive processes to be measured.
The material budget was varied in the simulation by $\pm 10 \%$ everywhere, and by $+ 50\%$ in the forward region only ($|\eta| > 1$). In both cases this did not modify the gap characteristics significantly. The maximum effect is for the largest $\Delta\eta$ bins in 2-arm events, and is still less than $10\%$. The effect was found to be negligible for triggering and event classification efficiencies.
In the region $|\eta| \leq 1$
the material budget is  known to better than $5\%$. 
In order to assess the sensitivity of the results to details of the detector-response simulation, the properties of the pseudorapidity distribution and gaps were also studied with the simulation output after stage $b$ (particle transport without detector response, using ideal hit positions). Only  negligible differences between ideal and real detectors were found.

The \mbor\ trigger covers the pseudorapidity range between $-3.7$ and $5.1$ except for a gap of $0.8$ units for $2.0 < \eta < 2.8$, which results in a small event loss. The proportion of events lost was estimated  by counting the number of events having tracks only in the corresponding interval on the opposite pseudorapidity side; the fraction loss of \mbor\ triggers was found to be below $10^{-3}$.

\subsection{Relative rate of single diffraction}

\begin{figure}[th!]
\begin{center}
\includegraphics[width=.82\textwidth]{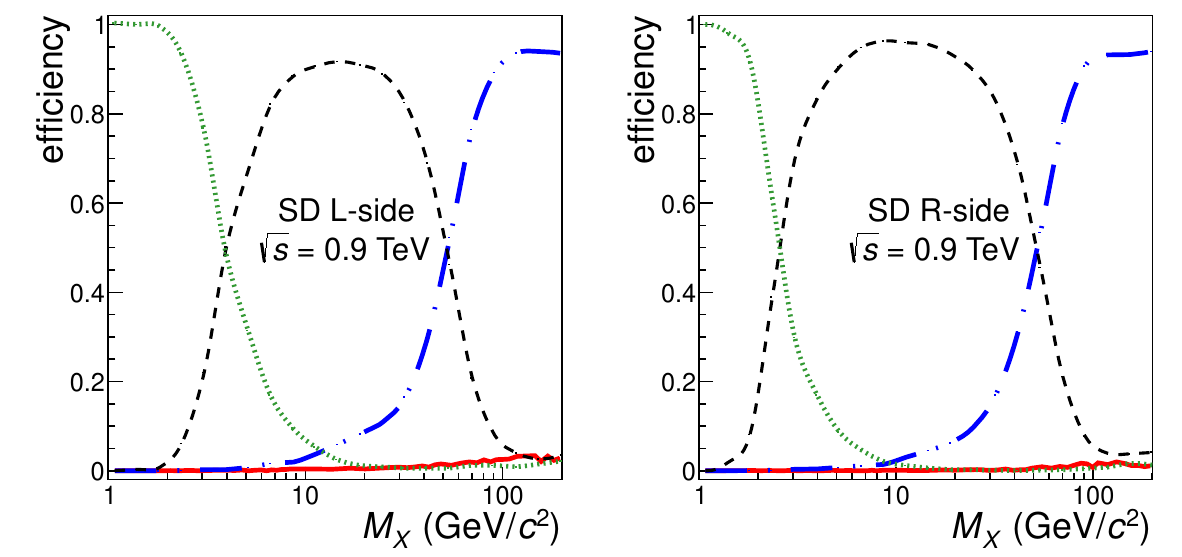}
\includegraphics[width=.82\textwidth]{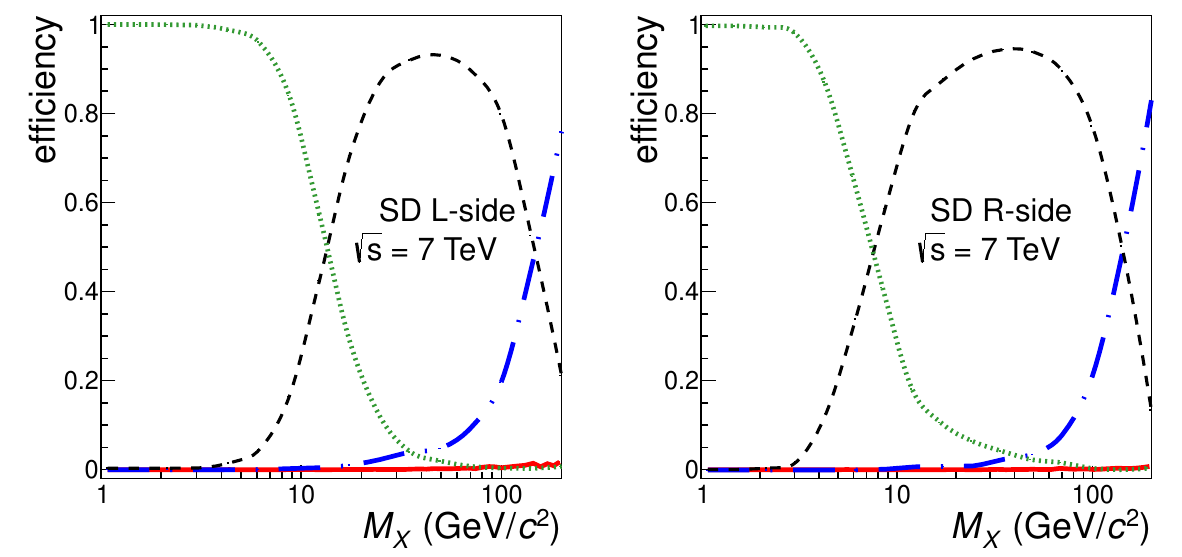}
\end{center}
\caption{
Detection efficiencies for SD events as a function of diffractive mass $M_X$ obtained by simulations with PYTHIA6, at $\sqrt{s}$ = 0.9~TeV (top), and 7~TeV (bottom).
L-side and R-side refer to the detector side at which SD occurred. Green dotted lines show the probability of not detecting the event at all.
Black dashed lines show the selection efficiency for an SD event on L(R)-side to be classified as the 1-arm-L(R) event. Blue dashed-dotted lines show the efficiency to be classified as a 2-arm event. Red continuous lines
show the (small) probability of L(R)-side single diffraction satisfying the 1-arm-R(L) selection, {\it i.e.} the opposite side condition.
}\label{Fig:fig9}
\end{figure}

The detection efficiencies for SD processes corresponding to the different
event classes, obtained with PYTHIA6 at $\sqrt{s}$ = 0.9~TeV and 7~TeV, are illustrated in Fig.~\ref{Fig:fig9}. For small diffractive masses, the produced particles have pseudorapidities close to that of the diffracted proton, therefore, such events are not detected. Increasing the
mass of the diffractive system broadens the distribution of emitted particles, and the triggered events are classified mostly as 1-arm-L or 1-arm-R class. Increasing the diffractive mass still further results in a substantial probability
of producing a particle in the hemisphere of the recoiling proton, and indeed for masses above $\sim 200$~GeV/$c^2$ such events end up mainly in the 2-arm class.
Because of multiplicity fluctuations and detection efficiencies, it is also possible for a SD event to be classified in the opposite
side 1-arm-R(L) class, albeit with a small probability (see Fig.~\ref{Fig:fig9}). Masses above $\sim 200$ GeV/$c^2$ end up mainly in the 2-arm class,
at all three energies. For this study, we have chosen $M_X = 200$ GeV/$c^2$ as the boundary between SD and NSD
events.
Changing the upper diffractive-mass limit in the definition of SD from $M_X = 200$~GeV/$c^2$ to $M_X = 50$~GeV/$c^2$ or $100$~GeV/$c^2$ at both
$\sqrt{s}$ = 0.9 and 7~TeV does not make a difference to the final results for the inelastic cross section, provided the data are corrected using the same model as that used
to parameterize the diffractive mass distribution. For example, at $\sqrt{s}$ = 0.9~TeV, if SD is defined for masses $M_X < 50$ GeV/$c^2$
($M_X < 100$ GeV/$c^2$), the measured SD cross section decreases by $20\%$ ($10\%$), which agrees with the predictions of the model~\cite{KP1} used for corrections.

\begin{table}[th!]
\begin{center}
\begin{tabular}{llccc}
\hline
  $\sqrt{s}$ (TeV) & Process  & 1-arm-L  & 1-arm-R  & 2-arm \\
\hline
           & SD L-side & $0.352 ^{+0.044} _{-0.014}$ &  $0.004 ^{+0.005} _{-0.003}$ &  $0.201 ^{+0.10} _{-0.05}$\\
  0.9    & SD R-side & $0.002 ^{+0.002} _{-0.001}$ &  $0.465 ^{+0.035} _{-0.031}$ &  $0.198 ^{+0.105} _{-0.054}$\\
           & NSD          & $0.012 \pm 0.004 $        &  $0.025 \pm 0.007$         &  $0.956 \pm 0.014$\\
\hline
            & SD L-side & $0.301 ^{+0.115} _{-0.021}$ & $0.002^{+0.003} _{-0.001}$  & $0.073 ^{+0.054} _{-0.027}$ \\
  2.76   & SD R-side & $0.002 ^{+0.002} _{-0.001}$ & $0.395 ^{+0.104} _{-0.011}$ & $0.087 ^{+0.071} _{-0.036}$\\
            & NSD          & $0.017 \pm 0.01$           & $0.026 \pm 0.008$          & $0.946 \pm 0.029$\\
\hline
            & SD L-side & $0.243 ^{+0.117} _{-0.029}$  & $0.0007^{+0.0010}_{-0.0006}$ & $0.041^{+0.032}_{-0.017}$\\
  7         & SD R-side & $0.0002^{+0.0003}_{-0.0002}$ & $0.333 ^{+0.121} _{-0.027} $ & $0.038^{+0.034} _{-0.019}$\\
            & NSD          & $0.013 \pm 0.003$  & $0.022 \pm 0.006 $              & $0.952\pm 0.014$\\
\hline
\end{tabular}
\end{center}
\caption{
Selection efficiencies at $\sqrt{s}$ = 0.9, 2.76 and 7~TeV for SD on the right and left sides and for NSD collisions to be classified as 1-arm-L(R) or 2-arm events. The errors listed are systematic errors; statistical errors are negligible.}
\label{Tab:tab1}
\end{table}

\begin{table}[th!]
\begin{center}
\begin{tabular}{llclcc}
\hline
  $\sqrt{s}$        & ratio         & ratio & side & \multicolumn{2}{c}{$\sigmaSD/\sigmainel$} \\
  (TeV)             & definition    &       &      &     per side                &  total   \\
\hline
   0.9              & 1-arm-L/2-arm & 0.0576 $\pm$ 0.0002 & L-side & 0.10 $\pm$ 0.02 & 0.21 $\pm$ 0.03 \\
                    & 1-arm-R/2-arm & 0.0906 $\pm$ 0.0003 & R-side & 0.11 $\pm$ 0.02 &                   \\
\hline
  2.76              & 1-arm-L/2-arm & 0.0543 $\pm$ 0.0004 & L-side & 0.09 $\pm$ 0.03 & $0.20^{+0.07}_{-0.08}$ \\
                    & 1-arm-R/2-arm & 0.0791 $\pm$ 0.0004 & R-side & $0.11^{+0.04}_{-0.05}$ &                   \\
\hline
  7                 & 1-arm-L/2-arm & 0.0458 $\pm$ 0.0001 & L-side & $0.10^{+0.02}_{-0.04}$ & $0.20^{+0.04}_{-0.07}$  \\
                    & 1-arm-R/2-arm & 0.0680 $\pm$ 0.0001 & R-side & $0.10^{+0.02}_{-0.03}$ &                   \\
\hline
\end{tabular}
\end{center}
\caption{
Measured  1-arm-L(R) to 2-arm ratios, and corresponding ratio of SD to INEL cross sections for three centre-of-mass energies.
Corrected ratios include corrections for detector acceptance, efficiency, beam background, electronics noise, and collision pileup. The
 total corresponds to the sum of SD from the L-side and the R-side. The errors shown are systematic uncertainties. In the 1-arm-L(R) to 2-arm
ratio, the uncertaities come from the estimate of the beam background. The uncertainty on the cross section ratio comes mainly from the
efficiency error listed in Table~\ref{Tab:tab1}. In all cases statistical errors are negligible.}
\label{Tab:tab2}
\end{table}
\begin{table}[th!]
\begin{center}
\begin{tabular}{lcccc}
\hline
$\sqrt{s}$ 	   & \mband\  	& \mbor\    &	\multicolumn{2}{c}{\mband/\mbor\ } \\
(TeV)          & ($\%$)     & ($\%$)    &   measured  & simulated \\
\hline
 0.9    & $76.3^{+2.2}_{-0.8}$      & $91.0^{+3.2}_{-1.0} $     & $0.8401 \pm 0.0004$    & $0.839^{+0.006}_{-0.008}$\\
2.76    & $76.0^{+5.2}_{-2.8}$      & $88.1^{+5.9}_{-3.5} $    & $0.8613 \pm 0.0006$    & $0.863^{+0.02}_{-0.03}$\\
7    & $74.2^{+5.0}_{-2.0}$      & $85.2^{+6.2}_{-3.0} $         & $0.8727 \pm 0.0001$    & $0.871 \pm 0.007 $\\
\hline
\end{tabular}
\end{center}
\caption{
\mband\  and \mbor\  trigger efficiencies obtained from the adjusted Monte Carlo simulations; comparison of  the measured and simulated trigger ratios \mband/\mbor\ at $\sqrt{s}$ = 0.9, 2.76 and 7~TeV. Errors shown are systematic uncertainties calculated in a similar way to that for  Table~\ref{Tab:tab1},
statistical errors are negligible.}
\label{Tab:tab3}
\end{table}

The efficiencies, obtained as the average between the adjusted PYTHIA6 and PHOJET values for three processes (L-side SD, R-side SD, and
NSD events) and for each event class are listed in Table~\ref{Tab:tab1} for the three energies under study. As these efficiencies depend on the adjustment of the event generators, and they are in turn used for the adjustment, one iteration was needed to reach the final values, as well as the final adjustment.
The systematic errors in Table~\ref{Tab:tab1} include an estimate of the uncertainty from the diffractive-mass distribution, and
take into account the difference of efficiencies between the two Monte Carlo generators and the uncertainty in the simulation
of the detector response. The uncertainty in the material budget is found to give a negligible contribution. In order to
estimate the systematic error due to the uncertainty on the diffractive-mass distribution, the dependence of the cross section
on diffractive mass from the model~\cite{KP1} was varied as described in Section~\ref{diffraction}, and, in addition,  the diffractive-mass distribution from the Donnachie-Landshoff model~\cite{DL} was used.

The raw numbers of events in the different classes were corrected for collision pileup by carrying out measurements for various runs with different average number of
 collisions per trigger.
The relative rates of SD events (cross-section ratios), Table~\ref{Tab:tab2},
 are calculated from the measured ratios of 1-arm-L(R) to 2-arm class events for a given DD fraction, which was adjusted as described above in this Section.
Even though the two sides of the detector are highly asymmetric and have significantly different acceptances, they give SD cross section
values that are consistent (Table~\ref{Tab:tab2}), which serves as a useful cross-check for the various corrections.

%


The SD fraction obtained at $\sqrt{s}$ = 0.9~TeV is found to be consistent with the UA5 measurement for p$\overline{\rm p}$ collisions~\cite{UA5}. The agreement with the UA5 result is much better if a $1/M_X$ diffractive-mass dependence is used for
our correction procedure, as was done for the UA5 measurements.


The \mband\ and \mbor\ trigger efficiencies (Table~\ref{Tab:tab3}) were obtained from the ALICE simulation, using the adjusted PYTHIA6 and PHOJET
event generators. 
In practice, for each assumption on the diffractive-mass distribution and for each fragmentation model, we determined together the diffractive fractions and the \mband\ and \mbor\ trigger efficiencies for detecting inelastic events.

An important cross-check of the simulation was obtained by comparing the measured and simulated ratios of the \mband\ to \mbor\
rates (Table~\ref{Tab:tab3}), which were found to be in agreement. The observed ratios were corrected for event pileup, using several runs with different
values of the average pileup probability.

\subsection{Relative rate of double diffraction}
DD events are defined as NSD events with a large pseudorapidity gap.
After adjustments, the Monte Carlo generators reproduce the measured gap width distributions (in the pseudorapidity range approximately $-3.7< \eta <5.1$) and the event ratios with reasonable accuracy. They may then be used to estimate the fraction of NSD events having a gap $\Delta\eta > 3$ in the full phase space, relative to all inelastic events. These fractions are given in Table~\ref{Tab:tab4}.
This $\Delta\eta$ value was chosen for the separation between DD and ND events in order to facilitate
comparison with lower energy data. Note that this DD relative rate includes a contribution from simulated events that were flagged by the event generators as ND. The fraction of such events is model-dependent and differs by a factor of two between PYTHIA6 and PHOJET. Up to $50\%$ of the DD events can be attributed to these ND-simulated events for $\Delta\eta > 3$.
\begin{table}[th!]
\begin{center}
\begin{tabular}{lc}
\hline
$\sqrt{s}$ (TeV)	& $\sigma_{\rm DD} / \sigma_{\rm INEL}$ \\
\hline
0.9	 & $0.11 \pm 0.03$ \\
2.76 & $0.12 \pm 0.05$ \\
7	 & $0.12^{+0.05}_{-0.04}$ \\
\hline
\end{tabular}
\end{center}
\caption{
Production ratios of DD with $\Delta\eta > 3$ to inelastic events, at $\sqrt{s}$ = 0.9, 2.76 and 7 TeV. The errors shown are systematic
uncertainties calculated in a similar way to that for  Table~\ref{Tab:tab1}, in all cases statistical errors are negligible.
}
\label{Tab:tab4}
\end{table}

\section{van der Meer scans}
\label{scans}
In order to determine the inelastic cross section, the luminosity has to be measured. The proton bunch current is obtained from induction signals in coils arranged around the beam pipe~\cite{HLC_BCNWG}, and van der Meer scans of the ALICE beam profiles are used to study the geometry of the beam interaction region.

The trigger condition \mband\ was used for this measurement. The rate $\mathrm{d}N/\mathrm{d}t$ for this trigger is given by
\begin{displaymath}
\frac{\mathrm{d}N}{\mathrm{d}t} = A \times \sigmainel \times \mathcal{L}.
\end{displaymath}
Here $A$ accounts for the acceptance and efficiency of the \mband\ trigger (determined in previous Section with adjusted simulations, Table~\ref{Tab:tab3}), $\sigmainel$ is the pp inelastic cross-section and $\mathcal{L}$ the luminosity. A simultaneous measurement of the LHC luminosity and the interaction rate determines the cross section $A \times \sigmainel$ for the \mband\ trigger (see Table~\ref{Tab:tab5}).

The luminosity for a single proton bunch pair colliding at zero crossing angle is given by
\begin{displaymath}
\mathcal{L} = fN_{1}N_{2}/h_{x}h_{y} ,
\end{displaymath}
where $f$ is the revolution frequency for the accelerator (11245.5 Hz for the LHC), $N_{1}$, $N_{2}$ the number of protons in each bunch, and $h_{x}$, $h_{y}$ the effective transverse widths of the interaction region. In practice, the effective width folds in small corrections for a non-zero crossing angle.

The parameters $h_{x}$ and $h_{y}$ are obtained from their respective rate-versus-displacement curves as the ratio of the area under the curve to the height at zero displacement. For Gaussian beam profiles
\begin{eqnarray*}
h_{x} = \sqrt{2\pi(\sigma_{1x}^{2}+\sigma_{2x}^{2})} \\
h_{y} = \sqrt{2\pi(\sigma_{1y}^{2}+\sigma_{2y}^{2})},
 \end{eqnarray*}
 where $\sigma_{ix}$, $\sigma_{iy}$ ($i = 1,2$ indexing the two beams) are the r.m.s. of the beams in the horizontal and vertical directions respectively. The van der Meer method is, however, valid for arbitrary beam shapes.

The VZERO detectors used to measure the \mband\ rate as a function of the horizontal and vertical displacement have almost constant acceptance during the scan, as the maximum displacement of the beams is 0.4~mm, to be compared to the distance of 0.9~m from the interaction point to the nearest VZERO array. The absolute displacement scale was calibrated by moving both beams in the same direction and measuring the corresponding vertex displacement with the SPD. This contributes with an uncertainty of 1.4\% to the $A \times \sigmainel$ measurement.

\begin{figure*}
\begin{center}
\includegraphics[width=.7\textwidth]{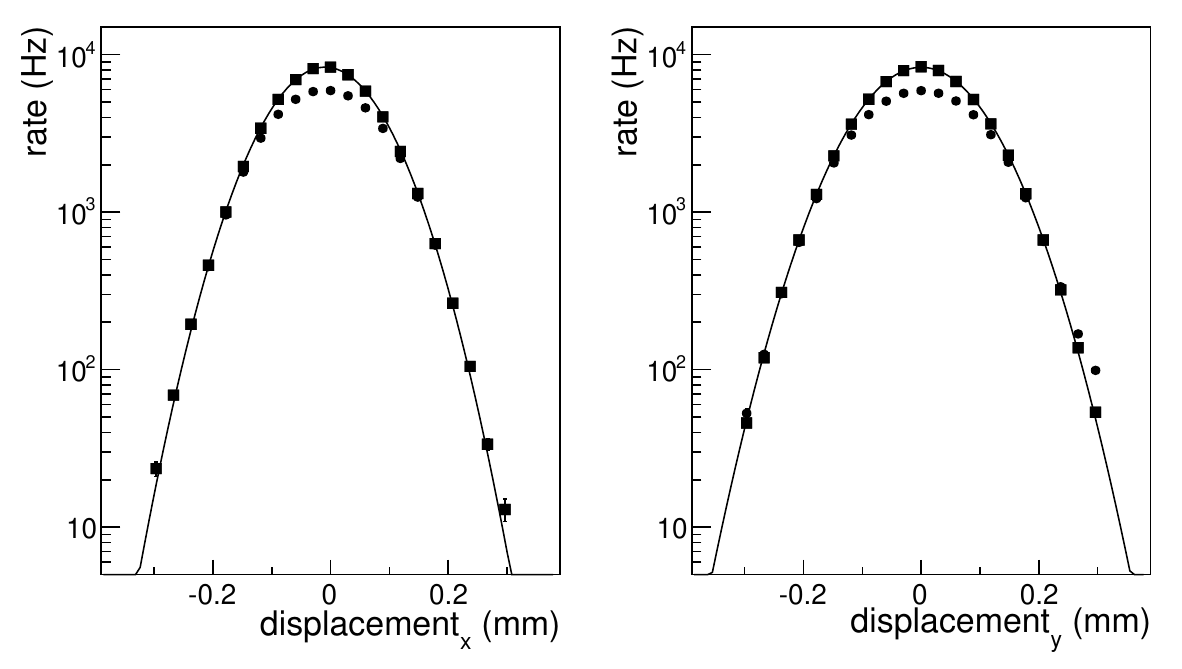}
\end{center}
\caption{
\mband\ trigger rates for horizontal (left) and vertical (right) relative displacements of the proton beams, for van der Meer scan II
performed at 7~TeV. Dots are raw trigger rates, squares are interaction rates after corrections discussed in the text. The lines
are to guide the eye. Only statistical errors are shown.
}\label{Fig:fig13}
\end{figure*}

Three separate scans were used for this analysis, as listed in Table~\ref{Tab:tab5}. Scans I and II, at 7 TeV, were performed at different times. They have significantly different beam parameters $\epsilon$ (emittance) and $\beta^{\ast}$ (interaction point amplitude function), where the transverse beam size $\sigma$ is related to these parameters as $\sigma^{2} = \epsilon\beta^{\ast}$. Scan II was repeated twice within a few minutes of each other using the same LHC fill. They show that the results of the measurement under near identical conditions are reproducible to within the statistical error of 0.3\%, so the average value was used in Table~\ref{Tab:tab5}. The displacement curves for scan II are shown in Fig.~\ref{Fig:fig13}.

\begin{table}[th!]
\begin{center}
\begin{tabular}{lcccccccc}
\hline
  Scan & $\sqrt{s}$  & colliding & crossing angle   & $\beta^{*}$ & $\mu$ at zero & $h_x/2\sqrt{\pi}$  & $h_y/2\sqrt{\pi}$ & $A \times \sigma_{\rm INEL}$   \\
           &  (TeV)  & bunches   & ($\mu$rad) &  (m)  & displacement &  ($\mu$m) &  ($\mu$m) &                      (mb) \\
\hline
I & 7    & 1  & 280 & 2   & 0.086  & 44  &  47 &  54.2 $\pm$ 2.9\\
II & 7    & 1  & 500 & 3.5 & 0.74   & 58  &  65 &  54.3 $\pm$ 1.9\\
III & 2.76 & 48 & 710 & 10  & 0.12   & 158 &  164&  47.7 $\pm$ 0.9\\
\hline
\end{tabular}
\end{center}
\caption{
For each van der Meer scan, centre-of-mass energy, number of colliding bunches, beam crossing angle, amplitude function at the interaction
point ($\beta^{*}$), average number of collisions per bunch crossing ($\mu$) at zero displacement, beam transverse size r.m.s. ($h_{x,y}/2\sqrt{\pi}$)
under the assumption of two identical Gaussian-shape beams, and measured minimum-bias cross section selected by \mband\ triggers with its systematic uncertainty.
\label{Tab:tab5}}
\end{table}

Several corrections were applied to the measurements to obtain the final cross sections and errors. The proton bunch intensities were corrected for ghost charge (protons circulating outside bunches)~\cite{LUMI:jeff} and for satellite charges (protons in subsidiary beam buckets). In addition, the following corrections were applied:

\begin{list}
{({\it \roman{romancounter}}$\thinspace$)}{\usecounter{romancounter}
\setlength{\rightmargin}{\leftmargin}}
\item background from beam-halo and beam--gas collisions: negligible for scans I and II, 30\% correction for scan III at maximum separation, leading however to only a 0.1\% correction for the cross section;
\item multiple collisions in a single bunch-crossing (pileup): 40\% correction to rate for scan II at zero displacement;
\item accidental triggers from noise or from trigger on two separate collisions: a maximum correction of $\sim 0.4$\% for scan II;
\item imperfect centering of beams: 0.7\% correction for scan II and negligible correction for scan III;
\item satellite collisions: these make a contribution to the rate for large $y$ displacements ({\it i.e.} 50\% rate correction at 300 $\mu$m displacement, giving however only a 0.1\% correction to the cross section;
\item luminosity decay during the scan: $\sim 1$\% correction.
\end{list}

For scan III the uncertainty on the bunch intensity was much lower
compared to scan II, so certain additional sources of uncertainty were
also investigated. These were: coupling between horizontal and
vertical displacements; variation of $\beta^{\ast}$ during the scan
resulting from beam--beam effects; further pulses in the VZERO
photomultipliers arising from ionization of the residual gas inside
the photomultiplier tube (after-pulses).\\ 
The pileup correction is the largest, however it results in a negligible contribution to the systematic uncertainty, because   it is a well-understood effect. 
This was checked by:
\begin{list}
{({\it \roman{romancounter}}$\thinspace$)}{\usecounter{romancounter}
\setlength{\rightmargin}{\leftmargin}}
\item checking the stability of the corrected \mband\ rate relative to the
rate of rare triggers, which are expected not to be influenced by
pileup;
\item  comparing resulting cross sections from data with different pileup fractions either in different scans, or in the same scan, but from different colliding bunches.
\end{list}
In all cases differences were negligible. Additionally, cross sections of "exclusive" triggers with conditions
such as the logical AND of V0-L(R) and not V0-R(L), which are strongly
affected by pileup, gave relative cross-sections stable in the full
range of pileup. This demonstrates that the pileup correction is well
understood \cite{Oyama}.\\
The contributions to the systematic uncertainty are listed in Table~\ref{Tab:tab6}.
Further details of these luminosity measurements are described in Ref.~\cite{Oyama}.


\begin{table}[th!]
\begin{center}
\begin{tabular}{lcc}
\hline
  $\sqrt{s}$ (TeV) -- scan   & 2.76 -- III   & 7 -- II \\
\hline
  Bunch current$^a$       & 0.53      & 3.1 \\
  Satellite charge$^b$      & 0.2        & 1  \\
\hline
  Bunch intensity (tot)$^c$       & 0.57      & 3.2 \\
\hline
Absolute displacement scale$^d$   & 1(h); 1(v)   & 1(h); 1(v) \\
Reproducibility             & 0.4          & 0.4     \\
Beam background           & 0.3        & Negl.   \\
$x$--$y$ displacement coupling & 0.6        & Negl. \\
Luminosity decay          & 0.5        & Negl.\\
$\beta^{\ast}$ variation during scan    & 0.4        & Negl. \\
VZERO after-pulse            & 0.2    & Negl.   \\
\hline
Experiment$^e$      & 1.75      & 1.5 \\
\hline
  Total                     & 1.84      & 3.5 \\
\hline
\end{tabular}
\end{center}
$^a$ bunch current uncertainty measured by the LHC BCNWG; includes ghost charge corrections\\
$^b$ satellite corrections to the beam current and to the trigger rate were evaluated by ALICE for scan II, and taken from \cite{LUMI:jeff} for scan III\\
$^c$ overall bunch intensity uncertainty\\
$^d$ separately for horizontal (h) and vertical (v) directions\\
$^e$ overall uncertainty  from the determination of the beam profiles
\caption{
Contributions to the systematic uncertainty in percentage of the minimum-bias cross section selected by the \mband\ trigger.
The beam intensity measurement was provided by the LHC Bunch Current Normalization Working Group (BCNWG)~\cite{HLC_BCNWG}.
\label{Tab:tab6}}
\end{table}

\section{Cross-section measurements}
\label{crosssection}
\subsection{Inelastic cross sections}
To obtain the inelastic cross section from the measurement of $A \times \sigma_{\rm INEL}$, discussed in Section~\ref{scans}, one must determine the factor $A$, which represents the \mband\ trigger acceptance and efficiency. The two previously introduced event generators, already adjusted for diffraction, together with the ALICE detector simulation, were used to determine this factor. The average values and their spread for the three energies are indicated in Table~\ref{Tab:tab3}.
The inelastic cross section is recalculated several times, using the two event generators and  four prescriptions for diffractive-mass distribution in SD process, as described in Section~\ref{diffraction}. For the two energies, where van der Meer scan measurements are available, the resulting inelastic pp cross sections are:
\begin{itemize}
   \item{$\sigma_{\rm INEL} = 62.8^{+2.4}_{-4.0}(model) \pm 1.2(lumi)$~mb at $\sqrt{s} = 2.76$~TeV;}
   \item{$\sigma_{\rm INEL} = 73.2^{+2.0}_{-4.6}(model) \pm 2.6(lumi)$~mb at $\sqrt{s} = 7$~TeV.}
\end{itemize}
The central values are the average of the two event generators with $M_X$ dependence given by model~\cite{KP1}. The first uncertainty, labeled as $model$, is determined from the upper and the lower results obtained using different assumptions. It also contains the influence of the variations in detector simulation described in Section~\ref{measurement}. However, it is dominated by the model assumptions (event generator, $M_X$ dependence).
For both energies the upper limit on the cross-section value is obtained with PHOJET and the $M_X$ dependence from model~\cite{KP1}, varied up by 50\% at the diffractive-mass threshold, and the lower limit with PYTHIA6 and the Donnachie--Landshoff parametrization~\cite{DL} of the $M_X$ dependence.
The second uncertainty, labeled as $lumi$, corresponds to the uncertainty in the determination of the
luminosity through van der Meer scans, as described in Section~\ref{scans}.

\begin{figure}[th!]
\begin{center}
\includegraphics[width=.47\textwidth]{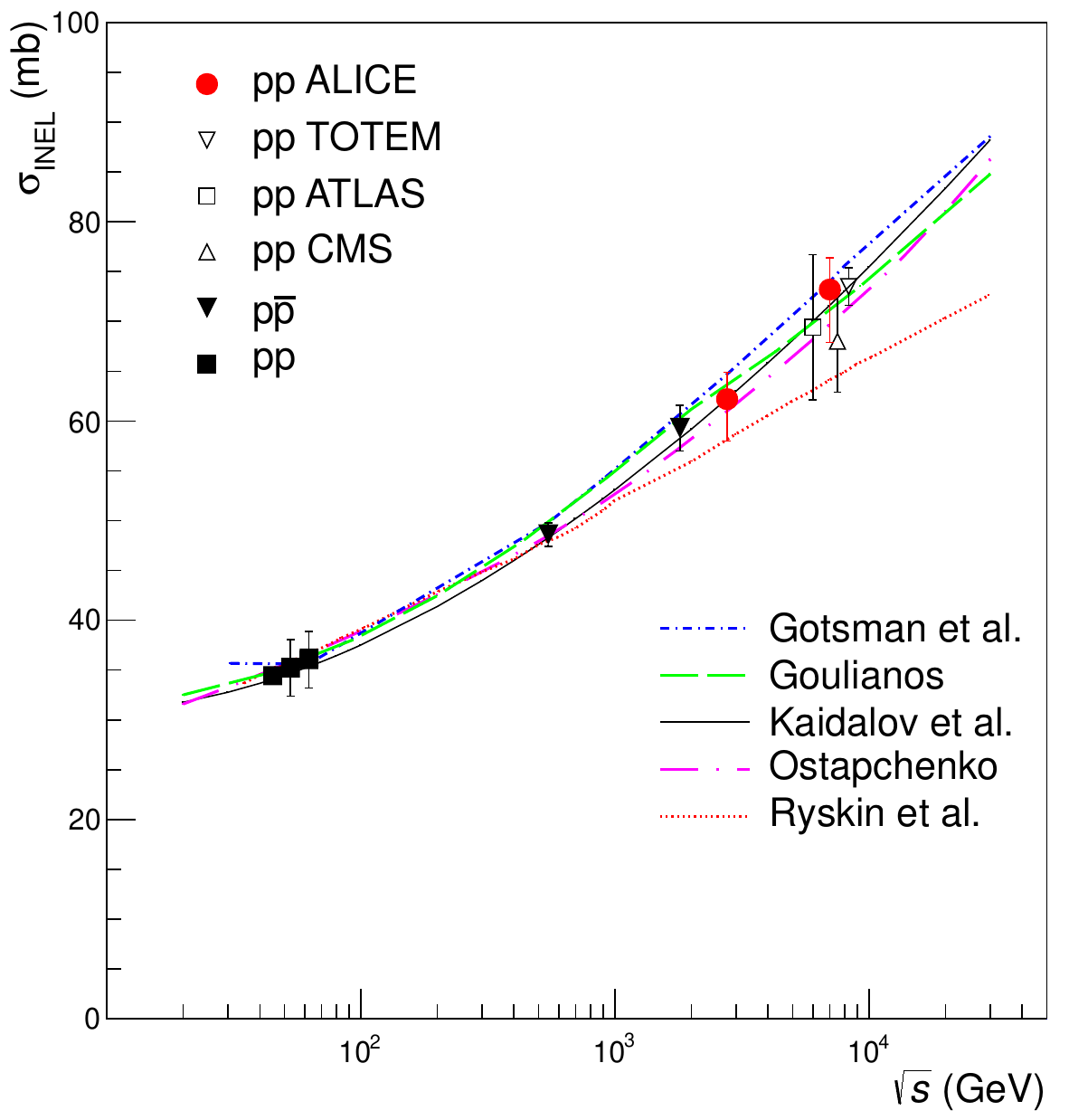}
\end{center}
\caption{
Inelastic cross sections as a function of centre-of-mass energy, in proton-proton or proton-antiproton collisions, compared with
predictions
\cite{GLM} (short dot-dashed blue line),
\cite{Goulianos} (dashed green line),
\cite{KP1} (solid black line),
\cite{Ostapchenko} (long dot-dashed pink line), and
\cite{KMR} (dotted red line).
LHC data are from ALICE [this publication], ATLAS \cite{ATLAS_InelXS},
CMS \cite{CMS_InelXS} and TOTEM \cite{Totem}.  Data points for ATLAS, CMS and TOTEM were
slightly displaced horizontally for visibility. Data from other experiments are taken from \cite{Inel_LED}.
}\label{Fig:fig14}
\end{figure}

The result at $\sqrt{s}$ = 7 TeV is consistent with measurements by ATLAS~\cite{ATLAS_InelXS}, CMS~\cite{CMS_InelXS}, and TOTEM~\cite{Totem} (Table~\ref{Tab:tab7}), albeit slightly higher
than the ATLAS and CMS values. A comparison of the ALICE results with other measurements at different energies and with models is shown
in Figure~\ref{Fig:fig14}. The LHC data favour slightly the higher prediction values.

\begin{table}[th!]
\begin{center}
\begin{tabular}{lll}
\hline
Experiment &  \multicolumn{1}{c}{$\sigma_{\rm INEL}$ (mb)} & \multicolumn{1}{c}{$\sigma_{\rm INEL}^{\xi > 5 \times 10^{-6}}$ (mb)} \\
\hline
ALICE       & $73.2^{+2.0}_{-4.6}(model) \pm 2.6(lumi)$ & $62.1^{+1.0}_{-0.9}(syst) \pm 2.2(lumi)$\\
ATLAS  & $69.4 \pm 6.9(model) \pm  2.4(exp) $ & $60.3 \pm 0.5(syst) \pm 2.1(lumi) $ \\
CMS      & $68.0 \pm 4.0(model) \pm  2.0(syst) \pm 2.4(lumi) $ & $60.2
\pm 0.2(stat) \pm 1.1(syst) \pm 2.4(lumi)$\\
TOTEM         & $73.5^{+1.8}_{-1.3}(syst) \pm 0.6(stat) $ & \\
\hline
\end{tabular}
\end{center}
\label{Tab:tab7}
\caption{
Inelastic cross section ($\sigma_{\rm INEL}$) measurements for pp collisions at $\sqrt{s}$ = 7~TeV at the LHC.
}
\end{table}

The ATLAS~\cite{ATLAS_InelXS} and CMS~\cite{CMS_InelXSwcut} collaborations published their results for
$\sigma_{\rm INEL}^{\xi > 5 \times 10^{-6}}$, which includes only
diffractive events with $M_X = \sqrt{\xi s} > 15.7$~GeV. These
measurements avoid the extrapolation to the low $M_X$ region, which is
the main source of systematic uncertainty on $\sigma_{\rm INEL}$. 
In our measurement of $\sigma_{\rm INEL}^{\xi > 5 \times 10^{-6}}$, about 40$\%$ of the uncertainty  comes from the $M_X$ dependence parameterization.
Table~\ref{Tab:tab7} also gives a comparison of inelastic
cross sections excluding low-mass diffraction, as measured by ALICE,
ATLAS and CMS. 
The results from three experiments are consistent within experimental uncertainties.

\subsection{Diffractive cross sections}
Combining the measurements of the inelastic cross section with the relative rates of diffractive processes, cross sections for
single ($M_X < 200$ GeV/$c^2$) and double ($\Delta\eta > 3$) diffraction were obtained:\\
\begin{itemize}
\item{$\sigma_{\rm SD} = 12.2^{+3.9}_{-5.3}(syst)$ mb and $\sigma_{\rm DD} = 7.8 \pm 3.2(syst)$ mb at $\sqrt{s} = 2.76$~TeV;}
\item{$\sigma_{\rm SD} = 14.9^{+3.4}_{-5.9}(syst)$ mb and $\sigma_{\rm DD} = 9.0 \pm 2.6(syst)$ mb at $\sqrt{s} = 7$~TeV.}
\end{itemize}
The inelastic cross section at $\sqrt{s}$ = 0.9~TeV was not measured by ALICE, instead, the value $\sigma_{\rm INEL} = 52.5^{+2.0}_{-3.3}$~mb was used, which includes the UA5 measurement~\cite{UA5xsec} and a re-analysis of the extrapolation to low diffractive masses~\cite{Poghosyan}.
Combining this value with the measured diffraction fraction (Table~\ref{Tab:tab2}), diffractive
cross sections were obtained at $\sqrt{s}$ = 0.9~TeV: $\sigma_{\rm SD} = 11.2^{+1.6}_{-2.1} (syst.)$~mb ($M_X < 200$~GeV/$c^2$) and
$\sigma_{\rm DD} = 5.6 \pm 2.0 (syst.)$~mb ($\Delta\eta > 3$). A summary of diffractive cross sections measured by ALICE is given in Table~\ref{Tab:tab8}.

\begin{table}[th!]
\begin{center}
\begin{tabular}{l c c}
\hline
$\sqrt{s}$ (TeV)  &	$\sigma_{\rm SD}$ (mb)  &	$\sigma_{\rm DD}$ (mb) \\
\hline
 0.9	& $11.2^{+1.6}_{-2.1}(syst)$   &	$5.6 \pm 2.0(syst)$ \\
 2.76	& $12.2^{+3.9}_{-5.3}(syst) \pm 0.2(lumi)$   &	$7.8 \pm 3.2(syst) \pm 0.2(lumi)$ \\
 7	    & $14.9^{+3.4}_{-5.9}(syst) \pm 0.5(lumi)$   &	$9.0 \pm 2.6(syst) \pm 0.3(lumi)$ \\
\hline
\end{tabular}
\end{center}
\label{Tab:tab8}
\caption{
Proton--proton diffractive cross sections measured by ALICE at $\sqrt{s}$ = 0.9, 2.76 and 7~TeV. Single diffraction is for $M_X < 200$~GeV/$c^2$ and double diffraction is for $\Delta\eta > 3$. The errors quoted are the total systematic uncertainties. Statistical errors are negligible.
}
\end{table}

\begin{figure}[th!]
\begin{center}
\includegraphics[width=.47\textwidth]{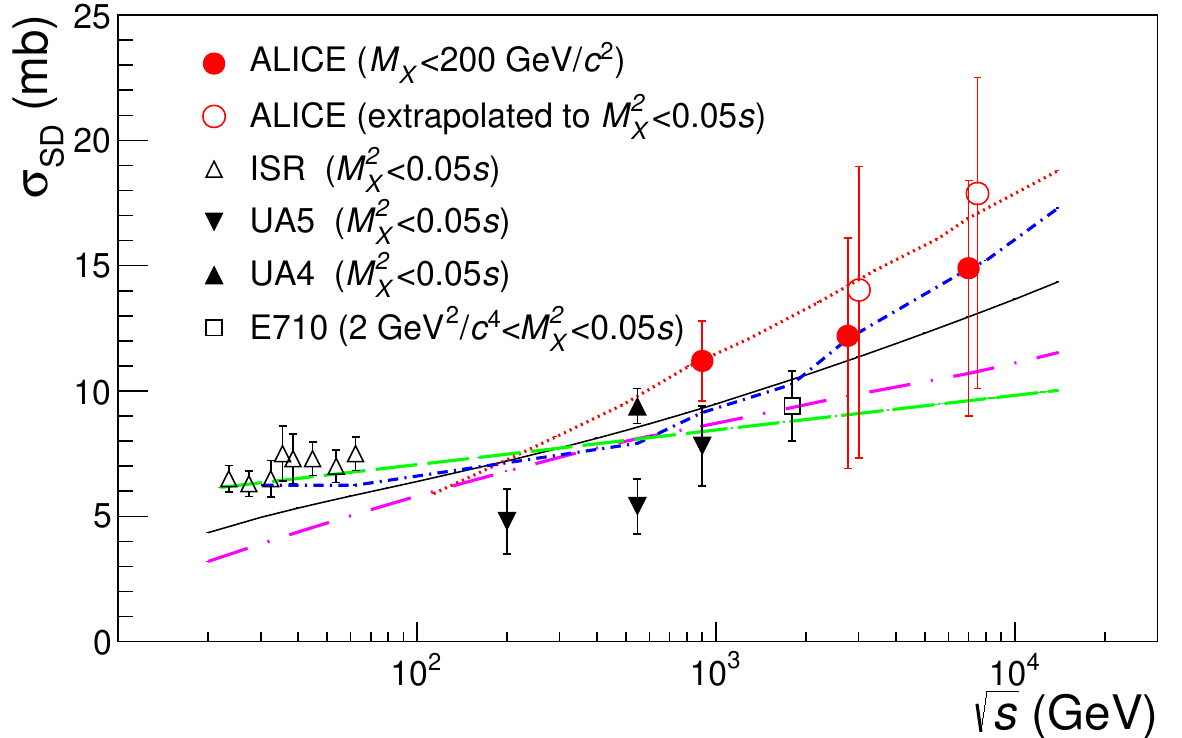}
\end{center}
\caption{
Single-diffractive cross section as a function of centre-of-mass energy. Data from other experiments are for $M_X^2 < 0.05s$ \cite{SD_LED}.
ALICE measured points are shown with full red circles, and, in order to compare with data from other experiments, were extrapolated to
$M_X^2 < 0.05s$ (open red circles), when needed. The predictions of theoretical models correspond to $M_X^2 < 0.05s$ and are defined as in
Fig.~\ref{Fig:fig14}.
}\label{Fig:fig15}
\end{figure}

\begin{figure}[th!]
\begin{center}
\includegraphics[width=.47\textwidth]{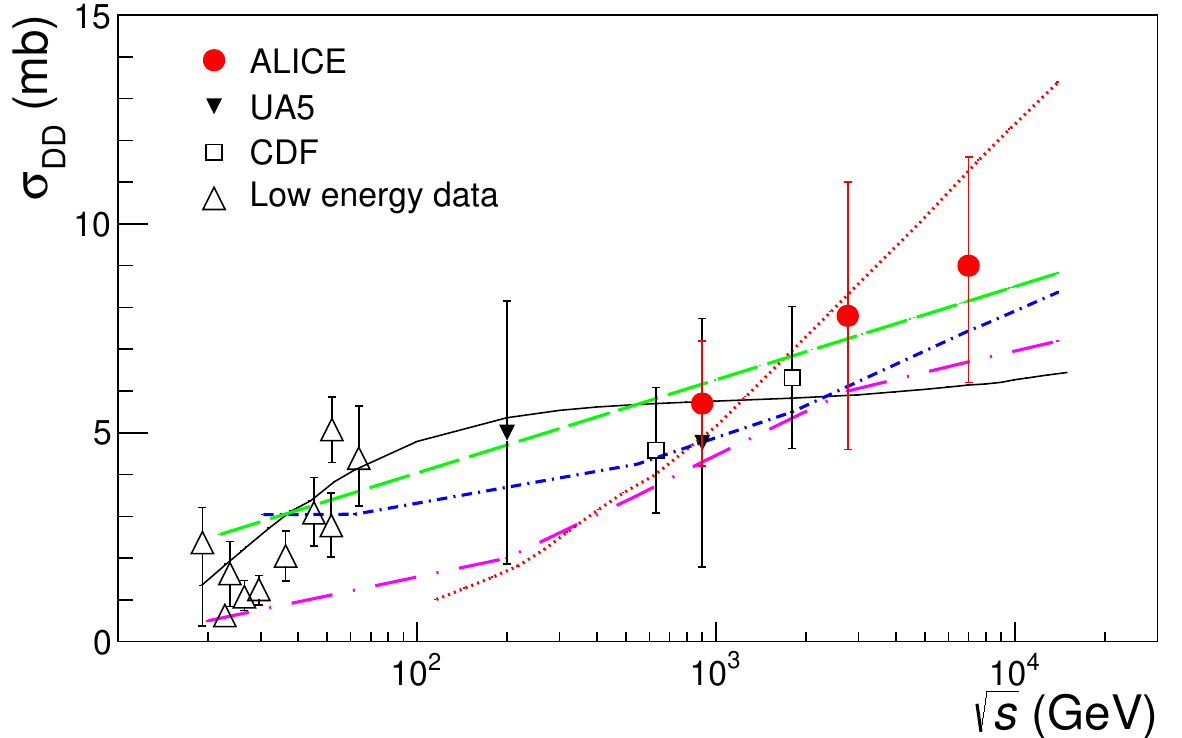}
\end{center}
\caption{
Double-diffractive cross section as a function of centre-of-mass energy. The theoretical model predictions represented as lines are
for $\Delta\eta > 3$ and are defined as in Fig.~\ref{Fig:fig14}. Data from other experiments are taken from \cite{DD_LED}.
}\label{Fig:fig16}
\end{figure}

A comparison of ALICE diffraction cross section measurements with data at previous colliders and with models is shown in Figs.~\ref{Fig:fig15} and \ref{Fig:fig16}. In order to facilitate comparison with  models, Fig.~\ref{Fig:fig15} also includes the SD cross section corrected (extrapolated) to the mass cut-off $M_X < \sqrt{0.05 s}$ ({\it i.e.} $\xi < 0.05$) at the energies 2.76 and 7~TeV.


A word of caution is needed concerning the comparison of data for SD and DD processes: results from different experiments are corrected in different ways, and also the definitions of SD and DD events are not unique. For example, the CDF collaboration\ \cite{affolder} defines DD events to be those with $\Delta \eta > 3$, as does this analysis, but in addition subtracts  non-diffractive events from their sample according to a model. In any case, within the large uncertainties, we  find agreement between ALICE measurements and data from
the CERN Sp$\overline{\rm p}$S collider and the Tevatron, as well as with the predictions of models \cite{KP1,GLM,KMR,Ostapchenko,Goulianos}.

\section{Conclusion}
A study of gaps in the pseudorapidity distributions of particles produced in pp collisions at the LHC was used to measure
the fraction of diffractive events in inelastic pp collisions at $\sqrt{s}$ = 0.9, 2.76 and 7~TeV. At $\sqrt{s}$ = 0.9~TeV,
the ALICE result on diffractive fractions is consistent with the UA5 data for p$\overline{\rm p}$ collisions.

The diffraction study made adjustments to the Monte Carlo generators used for evaluating trigger efficiencies.
The adjusted event-generator simulations together with the measurements
of the LHC luminosity with van der Meer scans were used to obtain the inelastic proton--proton
cross section at $\sqrt{s}$ = 2.76 and 7 TeV. The ALICE inelastic cross section result at $\sqrt{s}$ = 7 TeV is consistent with those
from ATLAS, CMS, and TOTEM.

Combining measured inelastic cross sections with diffraction relative rates, cross sections were obtained for single- and double-diffraction processes.

Cross section measurements were compared to other measurements at the LHC,
to lower energy data, and to predictions from current models \cite{KP1,GLM,KMR,Ostapchenko,Goulianos}, and
are found to be consistent with all of these, within present uncertainties.

\newenvironment{acknowledgement}{\relax}{\relax}
\begin{acknowledgement}
\section{Acknowledgements}
We are grateful to R. Ciesielski, E. Gotsman, K. Goulianos, V. Khoze,
G. Levin, S. Ostapchenko  and M. Ryskin for providing
us the numerical predictions of their models \cite{KMR, GLM,
Ostapchenko, Goulianos}.
\\
The ALICE collaboration would like to thank all its engineers and
technicians for their invaluable contributions to the construction of
the experiment and the CERN accelerator teams for the outstanding
performance of the LHC complex. 
The ALICE collaboration acknowledges the following funding agencies for their support in building and
running the ALICE detector:
Calouste Gulbenkian Foundation from Lisbon and Swiss Fonds Kidagan, Armenia;
Conselho Nacional de Desenvolvimento Cient\'{\i}fico e Tecnol\'{o}gico (CNPq), Financiadora de Estudos e Projetos (FINEP),
Funda\c{c}\~{a}o de Amparo \`{a} Pesquisa do Estado de S\~{a}o Paulo (FAPESP);
National Natural Science Foundation of China (NSFC), the Chinese Ministry of Education (CMOE)
and the Ministry of Science and Technology of China (MSTC);
Ministry of Education and Youth of the Czech Republic;
Danish Natural Science Research Council, the Carlsberg Foundation and the Danish National Research Foundation;
The European Research Council under the European Community's Seventh Framework Programme;
Helsinki Institute of Physics and the Academy of Finland;
French CNRS-IN2P3, the `Region Pays de Loire', `Region Alsace', `Region Auvergne' and CEA, France;
German BMBF and the Helmholtz Association;
General Secretariat for Research and Technology, Ministry of
Development, Greece;
Hungarian OTKA and National Office for Research and Technology (NKTH);
Department of Atomic Energy and Department of Science and Technology of the Government of India;
Istituto Nazionale di Fisica Nucleare (INFN) of Italy;
MEXT Grant-in-Aid for Specially Promoted Research, Ja\-pan;
Joint Institute for Nuclear Research, Dubna;
National Research Foundation of Korea (NRF);
CONACYT, DGAPA, M\'{e}xico, ALFA-EC and the HELEN Program (High-Energy physics Latin-American--European Network);
Stichting voor Fundamenteel Onderzoek der Materie (FOM) and the Nederlandse Organisatie voor Wetenschappelijk Onderzoek (NWO), Netherlands;
Research Council of Norway (NFR);
Polish Ministry of Science and Higher Education;
National Authority for Scientific Research - NASR (Autoritatea Na\c{t}ional\u{a} pentru Cercetare \c{S}tiin\c{t}ific\u{a} - ANCS);
Federal Agency of Science of the Ministry of Education and Science of Russian Federation, International Science and
Technology Center, Russian Academy of Sciences, Russian Federal Agency of Atomic Energy, Russian Federal Agency for Science and Innovations and CERN-INTAS;
Ministry of Education of Slovakia;
Department of Science and Technology, South Africa;
CIEMAT, EELA, Ministerio de Educaci\'{o}n y Ciencia of Spain, Xunta de Galicia (Conseller\'{\i}a de Educaci\'{o}n),
CEA\-DEN, Cubaenerg\'{\i}a, Cuba, and IAEA (International Atomic Energy Agency);
Swedish Research Council (VR) and Knut $\&$ Alice Wallenberg
Foundation (KAW);
Ukraine Ministry of Education and Science;
United Kingdom Science and Technology Facilities Council (STFC);
The United States Department of Energy, the United States National
Science Foundation, the State of Texas, and the State of Ohio.
\end{acknowledgement}
%
 
%
\newpage
%
\appendix
\section{The ALICE Collaboration}
\label{app:collab}

\begingroup
\small
\begin{flushleft}
B.~Abelev\Irefn{org1234}\And
J.~Adam\Irefn{org1274}\And
D.~Adamov\'{a}\Irefn{org1283}\And
A.M.~Adare\Irefn{org1260}\And
M.M.~Aggarwal\Irefn{org1157}\And
G.~Aglieri~Rinella\Irefn{org1192}\And
A.G.~Agocs\Irefn{org1143}\And
A.~Agostinelli\Irefn{org1132}\And
S.~Aguilar~Salazar\Irefn{org1247}\And
Z.~Ahammed\Irefn{org1225}
% ...
\renewcommand\labelenumi{\textsuperscript{\theenumi}~}
\section*{Affiliation notes}
\renewcommand\theenumi{\roman{enumi}}
\begin{Authlist}
\item \Adef{M.V.Lomonosov Moscow State University, D.V.Skobeltsyn Institute of Nuclear Physics, Moscow, Russia}Also at: M.V.Lomonosov Moscow State University, D.V.Skobeltsyn Institute of Nuclear Physics, Moscow, Russia
\item \Adef{University of Belgrade, Faculty of Physics and Institute of Nuclear Sciences, Belgrade, Serbia}Also at: University of Belgrade, Faculty of Physics and "Vin\v{c}a" Institute of Nuclear Sciences, Belgrade, Serbia
\end{Authlist}
\section*{Collaboration Institutes}
\renewcommand\theenumi{\arabic{enumi}~}
\begin{Authlist}
\item \Idef{org1279}Benem\'{e}rita Universidad Aut\'{o}noma de Puebla, Puebla, Mexico
\item \Idef{org1220}Bogolyubov Institute for Theoretical Physics, Kiev, Ukraine
\item \Idef{org1262}Budker Institute for Nuclear Physics, Novosibirsk, Russia
\item \Idef{org1292}California Polytechnic State University, San Luis Obispo, California, United States
\item \Idef{org1329}Central China Normal University, Wuhan, China
\item \Idef{org14939}Centre de Calcul de l'IN2P3, Villeurbanne, France
\item \Idef{org1197}Centro de Aplicaciones Tecnol\'{o}gicas y Desarrollo Nuclear (CEADEN), Havana, Cuba
\item \Idef{org1242}Centro de Investigaciones Energ\'{e}ticas Medioambientales y Tecnol\'{o}gicas (CIEMAT), Madrid, Spain
\item \Idef{org1244}Centro de Investigaci\'{o}n y de Estudios Avanzados (CINVESTAV), Mexico City and M\'{e}rida, Mexico
\item \Idef{org1335}Centro Fermi -- Centro Studi e Ricerche e Museo Storico della Fisica ``Enrico Fermi'', Rome, Italy
\item \Idef{org17347}Chicago State University, Chicago, United States
\item \Idef{org1288}Commissariat \`{a} l'Energie Atomique, IRFU, Saclay, France
\item \Idef{org1294}Departamento de F\'{\i}sica de Part\'{\i}culas and IGFAE, Universidad de Santiago de Compostela, Santiago de Compostela, Spain
\item \Idef{org1106}Department of Physics Aligarh Muslim University, Aligarh, India
\item \Idef{org1121}Department of Physics and Technology, University of Bergen, Bergen, Norway
\item \Idef{org1162}Department of Physics, Ohio State University, Columbus, Ohio, United States
\item \Idef{org1300}Department of Physics, Sejong University, Seoul, South Korea
\item \Idef{org1268}Department of Physics, University of Oslo, Oslo, Norway
\item \Idef{org1132}Dipartimento di Fisica dell'Universit\`{a} and Sezione INFN, Bologna, Italy
\item \Idef{org1315}Dipartimento di Fisica dell'Universit\`{a} and Sezione INFN, Trieste, Italy
\item \Idef{org1312}Dipartimento di Fisica dell'Universit\`{a} and Sezione INFN, Turin, Italy
\item \Idef{org1145}Dipartimento di Fisica dell'Universit\`{a} and Sezione INFN, Cagliari, Italy
\item \Idef{org1285}Dipartimento di Fisica dell'Universit\`{a} `La Sapienza' and Sezione INFN, Rome, Italy
\item \Idef{org1154}Dipartimento di Fisica e Astronomia dell'Universit\`{a} and Sezione INFN, Catania, Italy
\item \Idef{org1270}Dipartimento di Fisica dell'Universit\`{a} e Astronomia and Sezione INFN, Padova, Italy
\item \Idef{org1290}Dipartimento di Fisica `E.R.~Caianiello' dell'Universit\`{a} and Gruppo Collegato INFN, Salerno, Italy
\item \Idef{org1103}Dipartimento di Scienze e Innovazione Tecnologica dell'Universit\`{a} del Piemonte Orientale and Gruppo Collegato INFN, Alessandria, Italy
\item \Idef{org1114}Dipartimento Interateneo di Fisica `M.~Merlin' and Sezione INFN, Bari, Italy
\item \Idef{org1237}Division of Experimental High Energy Physics, University of Lund, Lund, Sweden
\item \Idef{org1192}European Organization for Nuclear Research (CERN), Geneva, Switzerland
\item \Idef{org1227}Fachhochschule K\"{o}ln, K\"{o}ln, Germany
\item \Idef{org1122}Faculty of Engineering, Bergen University College, Bergen, Norway
\item \Idef{org1136}Faculty of Mathematics, Physics and Informatics, Comenius University, Bratislava, Slovakia
\item \Idef{org1274}Faculty of Nuclear Sciences and Physical Engineering, Czech Technical University in Prague, Prague, Czech Republic
\item \Idef{org1229}Faculty of Science, P.J.~\v{S}af\'{a}rik University, Ko\v{s}ice, Slovakia
\item \Idef{org1184}Frankfurt Institute for Advanced Studies, Johann Wolfgang Goethe-Universit\"{a}t Frankfurt, Frankfurt, Germany
\item \Idef{org1215}Gangneung-Wonju National University, Gangneung, South Korea
\item \Idef{org1212}Helsinki Institute of Physics (HIP) and University of Jyv\"{a}skyl\"{a}, Jyv\"{a}skyl\"{a}, Finland
\item \Idef{org1203}Hiroshima University, Hiroshima, Japan
\item \Idef{org1254}Indian Institute of Technology Bombay (IIT), Mumbai, India
\item \Idef{org36378}Indian Institute of Technology Indore (IIT), Indore, India
\item \Idef{org1266}Institut de Physique Nucl\'{e}aire d'Orsay (IPNO), Universit\'{e} Paris-Sud, CNRS-IN2P3, Orsay, France
\item \Idef{org1277}Institute for High Energy Physics, Protvino, Russia
\item \Idef{org1249}Institute for Nuclear Research, Academy of Sciences, Moscow, Russia
\item \Idef{org1320}Nikhef, National Institute for Subatomic Physics and Institute for Subatomic Physics of Utrecht University, Utrecht, Netherlands
\item \Idef{org1250}Institute for Theoretical and Experimental Physics, Moscow, Russia
\item \Idef{org1230}Institute of Experimental Physics, Slovak Academy of Sciences, Ko\v{s}ice, Slovakia
\item \Idef{org1127}Institute of Physics, Bhubaneswar, India
\item \Idef{org1275}Institute of Physics, Academy of Sciences of the Czech Republic, Prague, Czech Republic
\item \Idef{org1139}Institute of Space Sciences (ISS), Bucharest, Romania
\item \Idef{org27399}Institut f\"{u}r Informatik, Johann Wolfgang Goethe-Universit\"{a}t Frankfurt, Frankfurt, Germany
\item \Idef{org1185}Institut f\"{u}r Kernphysik, Johann Wolfgang Goethe-Universit\"{a}t Frankfurt, Frankfurt, Germany
\item \Idef{org1177}Institut f\"{u}r Kernphysik, Technische Universit\"{a}t Darmstadt, Darmstadt, Germany
\item \Idef{org1256}Institut f\"{u}r Kernphysik, Westf\"{a}lische Wilhelms-Universit\"{a}t M\"{u}nster, M\"{u}nster, Germany
\item \Idef{org1246}Instituto de Ciencias Nucleares, Universidad Nacional Aut\'{o}noma de M\'{e}xico, Mexico City, Mexico
\item \Idef{org1247}Instituto de F\'{\i}sica, Universidad Nacional Aut\'{o}noma de M\'{e}xico, Mexico City, Mexico
\item \Idef{org23333}Institut of Theoretical Physics, University of Wroclaw
\item \Idef{org1308}Institut Pluridisciplinaire Hubert Curien (IPHC), Universit\'{e} de Strasbourg, CNRS-IN2P3, Strasbourg, France
\item \Idef{org1182}Joint Institute for Nuclear Research (JINR), Dubna, Russia
\item \Idef{org1143}KFKI Research Institute for Particle and Nuclear Physics, Hungarian Academy of Sciences, Budapest, Hungary
\item \Idef{org1199}Kirchhoff-Institut f\"{u}r Physik, Ruprecht-Karls-Universit\"{a}t Heidelberg, Heidelberg, Germany
\item \Idef{org20954}Korea Institute of Science and Technology Information, Daejeon, South Korea
\item \Idef{org1160}Laboratoire de Physique Corpusculaire (LPC), Clermont Universit\'{e}, Universit\'{e} Blaise Pascal, CNRS--IN2P3, Clermont-Ferrand, France
\item \Idef{org1194}Laboratoire de Physique Subatomique et de Cosmologie (LPSC), Universit\'{e} Joseph Fourier, CNRS-IN2P3, Institut Polytechnique de Grenoble, Grenoble, France
\item \Idef{org1187}Laboratori Nazionali di Frascati, INFN, Frascati, Italy
\item \Idef{org1232}Laboratori Nazionali di Legnaro, INFN, Legnaro, Italy
\item \Idef{org1125}Lawrence Berkeley National Laboratory, Berkeley, California, United States
\item \Idef{org1234}Lawrence Livermore National Laboratory, Livermore, California, United States
\item \Idef{org1251}Moscow Engineering Physics Institute, Moscow, Russia
\item \Idef{org1140}National Institute for Physics and Nuclear Engineering, Bucharest, Romania
\item \Idef{org1165}Niels Bohr Institute, University of Copenhagen, Copenhagen, Denmark
\item \Idef{org1109}Nikhef, National Institute for Subatomic Physics, Amsterdam, Netherlands
\item \Idef{org1283}Nuclear Physics Institute, Academy of Sciences of the Czech Republic, \v{R}e\v{z} u Prahy, Czech Republic
\item \Idef{org1264}Oak Ridge National Laboratory, Oak Ridge, Tennessee, United States
\item \Idef{org1189}Petersburg Nuclear Physics Institute, Gatchina, Russia
\item \Idef{org1170}Physics Department, Creighton University, Omaha, Nebraska, United States
\item \Idef{org1157}Physics Department, Panjab University, Chandigarh, India
\item \Idef{org1112}Physics Department, University of Athens, Athens, Greece
\item \Idef{org1152}Physics Department, University of Cape Town, iThemba LABS, Cape Town, South Africa
\item \Idef{org1209}Physics Department, University of Jammu, Jammu, India
\item \Idef{org1207}Physics Department, University of Rajasthan, Jaipur, India
\item \Idef{org1200}Physikalisches Institut, Ruprecht-Karls-Universit\"{a}t Heidelberg, Heidelberg, Germany
\item \Idef{org1325}Purdue University, West Lafayette, Indiana, United States
\item \Idef{org1281}Pusan National University, Pusan, South Korea
\item \Idef{org1176}Research Division and ExtreMe Matter Institute EMMI, GSI Helmholtzzentrum f\"ur Schwerionenforschung, Darmstadt, Germany
\item \Idef{org1334}Rudjer Bo\v{s}kovi\'{c} Institute, Zagreb, Croatia
\item \Idef{org1298}Russian Federal Nuclear Center (VNIIEF), Sarov, Russia
\item \Idef{org1252}Russian Research Centre Kurchatov Institute, Moscow, Russia
\item \Idef{org1224}Saha Institute of Nuclear Physics, Kolkata, India
\item \Idef{org1130}School of Physics and Astronomy, University of Birmingham, Birmingham, United Kingdom
\item \Idef{org1338}Secci\'{o}n F\'{\i}sica, Departamento de Ciencias, Pontificia Universidad Cat\'{o}lica del Per\'{u}, Lima, Peru
\item \Idef{org1316}Sezione INFN, Trieste, Italy
\item \Idef{org1271}Sezione INFN, Padova, Italy
\item \Idef{org1313}Sezione INFN, Turin, Italy
\item \Idef{org1286}Sezione INFN, Rome, Italy
\item \Idef{org1146}Sezione INFN, Cagliari, Italy
\item \Idef{org1133}Sezione INFN, Bologna, Italy
\item \Idef{org1115}Sezione INFN, Bari, Italy
\item \Idef{org1155}Sezione INFN, Catania, Italy
\item \Idef{org1322}Soltan Institute for Nuclear Studies, Warsaw, Poland
\item \Idef{org36377}Nuclear Physics Group, STFC Daresbury Laboratory, Daresbury, United Kingdom
\item \Idef{org1258}SUBATECH, Ecole des Mines de Nantes, Universit\'{e} de Nantes, CNRS-IN2P3, Nantes, France
\item \Idef{org1304}Technical University of Split FESB, Split, Croatia
\item \Idef{org1168}The Henryk Niewodniczanski Institute of Nuclear Physics, Polish Academy of Sciences, Cracow, Poland
\item \Idef{org17361}The University of Texas at Austin, Physics Department, Austin, TX, United States
\item \Idef{org1173}Universidad Aut\'{o}noma de Sinaloa, Culiac\'{a}n, Mexico
\item \Idef{org1296}Universidade de S\~{a}o Paulo (USP), S\~{a}o Paulo, Brazil
\item \Idef{org1149}Universidade Estadual de Campinas (UNICAMP), Campinas, Brazil
\item \Idef{org1239}Universit\'{e} de Lyon, Universit\'{e} Lyon 1, CNRS/IN2P3, IPN-Lyon, Villeurbanne, France
\item \Idef{org1205}University of Houston, Houston, Texas, United States
\item \Idef{org20371}University of Technology and Austrian Academy of Sciences, Vienna, Austria
\item \Idef{org1222}University of Tennessee, Knoxville, Tennessee, United States
\item \Idef{org1310}University of Tokyo, Tokyo, Japan
\item \Idef{org1318}University of Tsukuba, Tsukuba, Japan
\item \Idef{org21360}Eberhard Karls Universit\"{a}t T\"{u}bingen, T\"{u}bingen, Germany
\item \Idef{org1225}Variable Energy Cyclotron Centre, Kolkata, India
\item \Idef{org1306}V.~Fock Institute for Physics, St. Petersburg State University, St. Petersburg, Russia
\item \Idef{org1323}Warsaw University of Technology, Warsaw, Poland
\item \Idef{org1179}Wayne State University, Detroit, Michigan, United States
\item \Idef{org1260}Yale University, New Haven, Connecticut, United States
\item \Idef{org1332}Yerevan Physics Institute, Yerevan, Armenia
\item \Idef{org15649}Yildiz Technical University, Istanbul, Turkey
\item \Idef{org1301}Yonsei University, Seoul, South Korea
\item \Idef{org1327}Zentrum f\"{u}r Technologietransfer und Telekommunikation (ZTT), Fachhochschule Worms, Worms, Germany
\end{Authlist}
\endgroup

%
\end{document}